\documentclass[12pt]{article}
\usepackage{bm, amsmath, graphicx, enumerate, float, setspace, multirow, url, natbib}
\pdfoutput = 1

\begin{document}
	
	\bibliographystyle{apalike}
	
	\def\spacingset#1{\renewcommand{\baselinestretch}%
		{#1}\small\normalsize} \spacingset{1}

	\title{\bf A Novel Alternating Joint Longitudinal Model for Post-ICU Hemoglobin Prediction}
		\author{Gabriel Demuth, Ph.D. \thanks{
		This work was made possible in part by The Mayo Clinic Robert D. and Patricia E. Kern Center for the Science of Health Care Delivery}\hspace{.2cm}\\
	Department of Biomedical Statistics and Informatics, The Mayo Clinic\\
	Curtis Storlie, Ph.D. \\
	Department of Quantitiative Health Sciences, The Mayo Clinic\\
	Matthew A. Warner, M.D. \\
	Department of Anesthesiology, The Mayo Clinic\\
	Daryl J. Kor, M.D. \\
	Department of Anesthesiology, The Mayo Clinic\\
	Phillip J. Schulte, Ph.D. \\
	Department of Quantitative Health Sciences, The Mayo Clinic\\
	Andrew C. Hanson, M.S. \\
	Department of Quantitative Health Sciences, The Mayo Clinic\\
}
\maketitle
	
	\bigskip
	\begin{abstract}
		Anemia is common in patients post-ICU discharge. However, which patients will develop or recover from anemia remains unclear. Prediction of anemia in this population is complicated by hospital readmissions, which can have substantial impacts on hemoglobin levels due to surgery, blood transfusions, or being a proxy for severe illness. We therefore introduce a novel Bayesian joint longitudinal model for hemoglobin over time, which includes specific parametric effects for hospital admission and discharge. These effects themselves depend on a patient's hemoglobin at time of hospitalization; therefore hemoglobin at a given time is a function of that patient's complete history of admissions and discharges up until that time. However, because the effects of an admission or discharge do not depend on themselves, the model remains well defined. We validate our model on a retrospective cohort of 6,876 patients from the Rochester Epidemiology Project using cross-validation, and find it accurately estimates hemoglobin and predicts anemic status and hospital readmission in the 30 days post-discharge with AUCs of .82 and .72, respectively. 
	\end{abstract}
	
	\noindent%
	{\it Anemia, Recurrent Model, Bayesian Smoothing Spline} 
	\vfill
	
	\newpage
	\spacingset{1.1} 
	
	\section{Introduction} \label{introSect}
	
	Anemia occurs when a patient suffers from reduced oxygen-carrying capability, and is frequently associated with decreased quality of life \citep{ross2003effect}. Patients in the intensive care unit (ICU) very frequently develop anemia, with reported rates of 75\% to 95\% \citep{thomas2010anemia, corwin2004crit}. We use the \cite{AnemiaWHO} definition of anemia; hemoglobin $<$ 13 grams/deciliter for men, and $<$ 12 g/dL for non-pregnant women. 
	
	Although anemia is well defined, the ability to predict hemoglobin, and hence anemia, after discharge from the ICU remains limited. Previous efforts to understand the trajectory of hemoglobin post-discharge include \cite{warner2020prevalence}, who use hemoglobin scores at three month intervals post-discharge to assess recovery. Since the data is retrospective, however, only the closest observation to each three month timepoint could be used, leading to observations being ignored or moved somewhat in time. Additionally, this approach does not account for the effects of hospital admissions on hemoglobin. This is important because anemia is frequently caused by blood draws, surgery and other invasive procedures while in the hospital, and often treated while in hospital by blood transfusions \citep{thomas2010anemia}, all of which can cause large and rapid, but generally impermanent changes in hemoglobin levels. Because of these effects, modeling the effects of hospitalization and recovery is important for capturing the dynamics of hemoglobin following ICU discharge. 
	
	To account for these challenges, we represent each patient's hemoglobin via a patient-specific random effects model, which consists of an overall trend representing the patient's hemoglobin absent any hospitalizations, and additive random effects for each hospitalization. We model both the trend and hospitalization effects using Bayesian Smoothing Spline ANOVA (BSS-ANOVA), which is a basis expansion derived from a Gaussian Process \citep{storlie2013methods, storlie2015calibration}. The BSS basis is convenient for Bayesian applications, because it implies that the regression coefficients are independent and normally distributed, with a common variance parameter. Because higher order basis elements are higher frequency and lower magnitude, the common variance protects against overfitting by shrinking the more sensitive high order elements towards zero.
	
	The trend plus hospitalization random effects approach models hemoglobin conditioned on known hospitalization times. However, any predictive use of the model must account for the potential effects of unknown future hospitalizations. Because hospital admission and discharge are likely to be themselves driven by hemoglobin levels, we employ a joint longitudinal model (JLM), which represents data where a time to event process is correlated longitudinal data \citep{tsiatis2004joint, rizopoulos2010jm}. Typically JLMs use individual-specific random effects to model the ``true" value of longitudinal variables, and specify the event process conditional on these true values \citep{tsiatis2004joint}. We broadly follow this pattern, specifying event distributions conditional on the random effects model for hemoglobin described above. 
	
	However, we make two departures from standard JLMs. A hospitalization consists of an admission and a discharge; a patient at home is only exposed to admission risk and a hospitalized patient is only exposed to discharge risk. Because being hospitalized and being at home are not absorbing states, each process is recurrent. Since patients switch between states this is what \cite{hougaard1999multi} terms an alternating model. While there is an existing literature on alternating and recurrent models, most of these only use baseline covariates that are constant in time. A notable exception to this is \cite{han2007parametric}, which extends the latent class approach of \cite{lin2002latent} to a recurrent, but not alternating, model. In this approach each individual belongs to one of a relatively small number of latent classes, which partially determines the trajectory of their longitudinal variables. In this approach the event process depends only on the class membership and not on the longitudinal variable itself. 
	
	By contrast, we allow both admission and discharge processes to depend directly on the true hemoglobin value. So far as we are aware, this work is the first to present a recurrent or alternating JLM with the event processes depending directly on the longitudinal variable. Further, as mentioned previously, each admission and discharge have associated random effects that directly impact patients' hemoglobin going forwards in time. Thus our JLM allows two-way feedback between the longitudinal and event processes; hemoglobin impacts the admission and discharge risks, which alters future hemoglobin, which effects the risk of subsequent admissions. This appears to be a substantial innovation in the field of JLMs, and allows for a much richer and more dynamic representation of the interplay between longitudinal and event data. 
	
	The rest of the paper is organized as follows. Section \ref{dataSect} describes the data used for the study, including collection and basic demographics. Section \ref{modelSect} describes the features of our model, broken down into trend (Section \ref{trendSect}), hospitalization (Section \ref{hospSect}), admission and discharge (Section \ref{eventSect}) models, as well as methods for fitting and prediction (Section \ref{predSect}). Finally, Section \ref{resultsSect} demonstrates the efficacy of our model in predicting hemoglobin levels, anemic status, and future hospitalization, while Section \ref{concludeSect} concludes by discussing the overall model, its successes, limitations, and areas of future interest. 
	
	We use Greek letters for random variables and Arabic characters for observed data and hyperparameters, values for which are given in Table \ref{hyperTab} in the Supplemental Materials (SM). Boldface indicates a vector, matrices are written in script, e.g. $\mathcal{X}_i$. The subscript $k$ and superscript $\prime$ are used throughout for arbitrary indexing or indicating a specific value, with meaning defined in each context. 
	
	\section{Post-ICU Data} \label{dataSect}
	The data for this study are derived from the Rochester Epidemiology Project (REP) database, a medical records linkage system which contains comprehensive population health information for residents in southeastern Minnesota. Specifically, we included data from 6,876 adult ($\geq$ 19 years of age) residents of Olmsted County, Minnesota who had been hospitalized at any of the qualifying medical centers and admitted to an intensive care unit (ICU) at least once between January 1, 2010 and January 16, 2019. This ICU visit is considered their index hospitalization, as only patients with such a hospitalization are included in the data, and observation times, ages, etc are all calculated relative to this encounter. Inclusion and exclusion criteria have been described previously \citep{warner2020prevalence}, but briefly, patients must have survived their index hospitalization, and had at least 2 independent hemoglobin observations obtained during the index hospital admission.
	
	Basic demographic data at time of the index discharge is available for all patients; women comprise 45\% of patients, with an average age of 65.6 years (sd = 19.6, min = 19, max = 105.7 years). Men constitute 55\% of the study population, with an average age of 63.3 years (sd = 18.5, min = 19, max = 103.7 years). 
	
	In order to better estimate each patient's random effects, we use a full year's data (365 days) before and after discharge from the index hospitalization. We therefore consider day 0 to be day of index discharge, patients enter the data at day -365, and remain in the data until day 365. We require patients to be at least 19 years old at index discharge, which ensures no pediatric data is included in the prior year. Patients average 23.6 hemoglobin observations over the course of the study period; however there is substantial variation in observation count between patients (sd = 27.4, min = 2, max = 455). About 77\% of observations fall below WHO definition of anemia, which is consistent with rates found by \cite{thomas2010anemia, corwin2004crit}. Patients experienced an average of 2.4 hospitalizations including the index encounter (sd = 2.1, min = 1, max = 35 hospitalizations), with an average duration of 6.2 days (sd = 8.84, min = .02, max = 241 days). Because we do not include death as an outcome in this model, we treat patient mortality as a censoring event. Censoring due to mortality is common; around 39\% of patients are censored for death or other reasons in the year following index discharge, with an average post-discharge survival of 310.3 days (sd = 121.1, min = 0, max = 365 days).
	
\begin{figure}[h!]
		\centering
		\includegraphics[width=\textwidth]{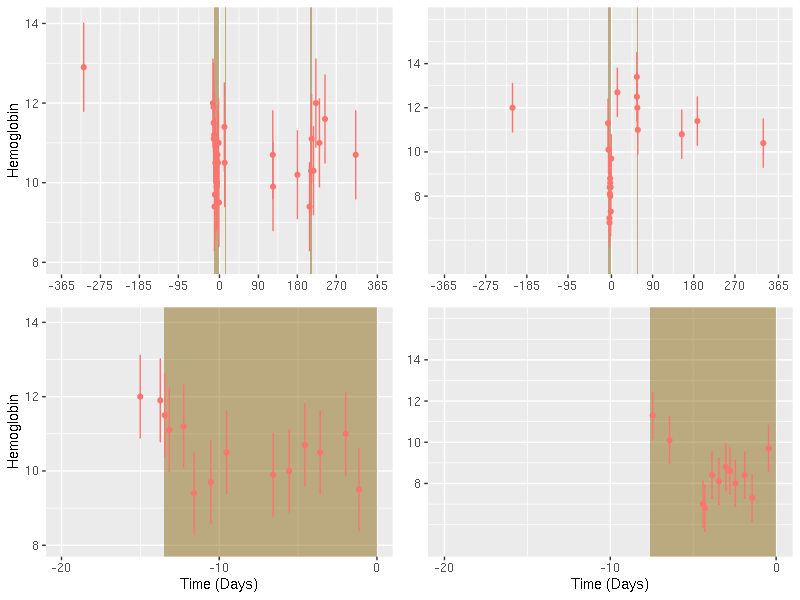}
		\caption{\label{rawHbFig} \footnotesize Observed hemoglobin values for two sample patients (left, right columns), across the full two years (top row) and detail of their index encounters (bottom row). The shaded rectangles indicate periods of hospitalization. Observed hemoglobin values are plotted with $+/-2$ standard deviation error bars, sd is estimated by the posterior median of the fit model. }
	\end{figure}

	Hospitalization plays a key role in the data. Figure \ref{rawHbFig} shows observed hemoglobin and hospitalization times for two patients. 
	
	Figure \ref{rawHbFig} shows that observations are irregular and tend to be clustered within hospitalization times. Fully 58\% of observations are made during a hospital visit. This alone makes simply removing hemoglobin observations made during hospitalization unrealistic, there simply isn't enough data left to accurately estimate a model or have a meaningful trend. Further, hospitalizations are associated with rapid changes in hemoglobin, with potentially longer recovery times after discharge, which will impact later non-hospitalized observations. Any model that did not incorporate hospitalization and recovery would struggle to handle these effects. 
	
	Hospitalization itself is likely a proxy variable for a wide variety of underlying phenomena that drive these rapid hemoglobin changes, such as some critical illness, surgeries or blood transfusions. While ideally these could be modeled directly, the potentially vast number of different treatments and illness and limited data availability renders this detailed approach unrealistic in practice. Hospitalization therefore serves as an important indicator that a patient's hemoglobin levels are likely to change in the near future, and as a proxy for a variety of other, more complex medical phenomena. Hospital readmission is also an interesting outcome in its own right, with a substantial literature devoted to attempting to predict it; see \cite{kansagara2011risk}, \cite{artetxe2018predictive} and \cite{mahmoudi2020use} for three systematic reviews of the subject. 
	
	\section{Modeling Hemoglobin and Readmission} \label{modelSect}
	We present the model as follows. First we develop a simple trend model, which represents how a patient's hemoglobin would evolve over time if no hospitalizations were to occur. We then introduce additive effects for hospital admissions and rate of recovery after discharge. Next we define time-to-event models for readmission and discharge conditional on hemoglobin, which completes the joint model for the event data and longitudinal variables. Finally, we briefly describe model fitting, as well as how to use a fit model to generate predictions of future hemoglobin and hospitalization.  
	
	\subsection{Trend Model} \label{trendSect}
	Although hospitalization is important, and will be included in the final model, we first consider estimation of a patient's hemoglobin trend absent hospitalization.
	
	Let $H_i(t_{i,s})$ denote the observed hemoglobin for patient $i, i = 1, \dots, n$ at his/her individually specific observation times $t_{i,s}, s = 1, \dots, n_i$. Let $\bm{H}_i$ be the $n_i$-vector of all hemoglobin values for patient $i$. Let $\mathcal{X}_i$ be an $n_i \times p$ (including the intercept) BSS-ANOVA expansion, the details of which are given below, of time for patient $i$, and $\bm{\alpha}_i$ be a vector of individual-specific random effects. Then we assume
	
	
	\begin{equation} \label{responseModel}
		\begin{split}
			&\bm{H}_i | \bm{T}_i, \sigma^2 \overset{ind}{\sim} N(\bm{T}_i, \sigma^2 I), i = 1, \dots, n,  \\
			&\bm{T}_i | \bm{\alpha}_i = \mathcal{X}_i \bm{\alpha}_i, \\
			& \sigma^2 \sim inv. Gamma(a_\sigma, b_\sigma), \\
		\end{split}
	\end{equation}
	
	where $\sigma^2$ is an unknown variance component representing measurement error. We specify the response model in terms of $\bm{T}_i$, as this quantity is important throughout the model, and is modified in Section \ref{hospSect} to account for hospitalizations.
	
	
	The BSS-ANOVA basis is a spline-like basis expansion derived from the covariance function of a Gaussian Process (GP), see \cite{storlie2013methods, storlie2015calibration} and \cite{reich2009variable} for details. Because it is based on a covariance function with domain $[0,1]$, BSS-ANOVA requires its inputs to be scaled to $[0,1]$. The $[s,k]$th element of $\mathcal{X}_i$ is given by $\tilde{\phi}_{k - 1}(t_{i,s}) = \sqrt{\pi_{k - 1}} \phi_{k - 1}(t_{i,s}), s = 1, \dots, n_i, k = 1, \dots p$, where $\phi_{k - 1}$ is the $k - 1$st eigenfunction of the GP's Karhunen-Lo\`eve representation, and $\pi_{k - 1}$ is the corresponding eigenvalue. Indexing the eigenfunctions and eigenvalues by $k - 1$ includes the intercept, which has eigenfunction $\phi_0(t) = 1$ and eigenvalue $\pi_0 = 1$. As $k$ increases, the eigenvalues approach zero, while the eigenfunctions increase in frequency.
	
	The GP origin of the BSS basis implies that the $\alpha_{i,k}$ are independent and normally distributed, the intercept has separate variance $\tau^2_0$ and all other $\alpha_{i,k}, k > 1$ have common variance $\tau^2$. This common variance acts to shrink the higher frequency basis elements (due to the decreasing magnitude of the eigenvalue multipliers) and thus protects against overfitting, much like the elastic net or ridge regression.  Therefore, although the covariance function has an infinite number of eigenfunctions and eigenvalues, for modeling purposes only a fairly small number have any practical impact. Thus, we may include only a moderate number (e.g. $p = 10$) with no noticeable effect on estimation. If in doubt, the model may be refit using a greater number of basis functions (say $p = 20$) to verify that additional elements do not impact the final estimated fit significantly.
	
	In the original BSS-ANOVA formulation, all the regression coefficients have prior mean zero. However, a patient's hemoglobin is likely to be influenced by their sex, age, disease diagnoses and other baseline variables. Let $\bm{z}_{i,0}$ be a vector of baseline variables for patient $i$, such including factors such as age, sex or diagnosed illness that may influence the intercept, and let
	
	
	\begin{equation} \label{alpha0Prior}
		\alpha_{i,0} | \bm{\gamma}_0, \tau^2_0 \overset{ind}{\sim} N(\bm{z}_{i,0}^T \bm{\gamma}_0, \tau^2_0), i = 1, \dots, n,
	\end{equation}
	
	\noindent for unknown population mean vector $\bm{\gamma}_0$. Similarly, let $\bm{z}_i$ be a vector of potentially different baseline variables for patient $i$ that effect the non-intercept elements of $\bm{\alpha}_i$, and assume 
	
	\begin{equation} \label{alphaPrior}
		\alpha_{i,k} | \bm{\gamma}_k, \tau^2 \overset{ind}{\sim} N(\sqrt{\pi}_{k - 1} \bm{z}_i^T \bm{\gamma}_k, \tau^2), k = 2, \dots, p, \; i = 1, \dots, n, 
	\end{equation}
	
	\noindent where $\bm{\gamma}_k$ is an unknown mean vector for the $k$th component. All variables included in $\bm{z}_{i,0}$ are described in Table \ref{covariateTab} in the SM; we use an intercept-only model in (\ref{alphaPrior}). We use convenient conjugate priors for $\bm{\gamma}_0, \bm{\gamma}$ and the variance parameters $\sigma^2_{\gamma_0}$, $\sigma^2_\gamma$.
	
	\begin{equation} \label{gammaPrior}
		\begin{split}
			&\bm{\gamma}_0 | \sigma^2_{\gamma_0} \sim N(\bm{0}, \sigma^2_{\gamma_0} I), \\
			&\bm{\gamma}_k | \sigma^2_\gamma \overset{ind}{\sim} N(\bm{0}, \pi_k \sigma^2_\gamma I), k = 2, \dots, p, \\
			&\tau^2_0 \sim inv. Gamma(a_{\tau_0}, b_{\tau_0}), \\
			&\tau^2 \sim inv. Gamma(a_\tau, b_\tau), \\
			&\sigma^2_{\gamma_0} \sim inv. Gamma(a_{\gamma_0}, b_{\gamma_0}), \\
			&\sigma^2_\gamma \sim inv. Gamma(a_\gamma, b_\gamma).
		\end{split}
	\end{equation} 
	
	The structure of (\ref{alpha0Prior}), (\ref{alphaPrior}) and (\ref{gammaPrior}) creates two levels of shrinkage. Each $\alpha_{i,k}, k > 1$ has a prior mean, which is shrunk towards zero as $k$ increases and $\sqrt{\pi_{k-1}}$ decreases. This shrinkage is enforced in the hyperprior by scaling the common variance $\sigma^2_\gamma$ by $\pi_{k-1}$, which prevents the means of the $\alpha_{i,k}$ from growing arbitrarily through inflating $\bm{\gamma}_k$. This structure allows incorporation of prior information, but safeguards strongly against overfitting. 
	
	
	The error variance, $\sigma^2$, is a major driver of model behavior, since it determines the penalty for lack of fit in (\ref{responseModel}). We therefore use a very informative prior, $a_\sigma = 400001, b_\sigma = 100000$, which strongly enforces a prior mean of $0.25$. This was chosen based on the understood accuracy of hemoglobin measurements (Dr. Matthew Warner, personal communication), and forces the model to respect the known degree of measurement error. 
	
	\subsection{Hospitalization and Recovery Effects} \label{hospSect}
	The model described in (\ref{responseModel}) heavily penalizes rapid changes in hemoglobin, which is appropriate since hemoglobin generally changes slowly. However, surgeries, blood transfusions, and other medical complications can cause sudden and large changes in hemoglobin levels. Since these events are likely to occur during hospitalizations, we create a submodel that allows for departure from the trend $\mathcal{X}_i \bm{\alpha}_i$ during hospitalization, followed by a continuous return to the trend after discharge. 
	
	Recalling Figure \ref{rawHbFig}, it is clear that any hospitalization model must allow for multiple hospitalizations and discharges; the patient on the right had three separate hospitalizations. It is also likely that the effects of hospitalization are both cumulative, in that being admitted a second time does not automatically erase the effects of the first, and transient, in that as time passes since a hospitalization, a patient should tend back towards their tren,d $\mathcal{X}_i \bm{\alpha}_i$. Finally, the hospitalization model must be locally flexible, since hemoglobin can change fairly rapidly during a hospital encounter, due to surgeries and transfusions. 
	
	Suppose patient $i$ experiences $J_i$ hospitalizations during the study period, with hospitalization $j$ lasting from admission time $a_{i,j}$ to discharge time $d_{i,j}$, with $a_{i,1} < d_{i,1} < a_{i,2} < \dots < d_{i,J_i}$. We then extend (\ref{responseModel}) with additive terms for each hospitalization,
	
	\begin{equation} \label{readmitResponse}
		\bm{T}_i | \bm{\alpha}_i, \lambda_{i,j}, \bm{\beta}_{i,j}, j = 1, \dots J_i = \mathcal{X}_i \bm{\alpha}_i + \sum_{j = 1}^{J_i} \mathcal{C}_{i,j} \bm{\beta}_{i,j}
	\end{equation}
	
	\noindent with $\mathcal{X}_i, \bm{\alpha}_i$ as before. Here $\bm{\beta}_{i,j}$ is a length $b$ random vector describing the impact of the $j$th hospitalization on patient $i$, and $\mathcal{C}_{i,j}$ is the $n_i \times b$ basis expansion for that hospitalization, defined as follows. Let $\tilde{\phi}_k = \sqrt{\pi_k} \phi_k$, with $\pi_k, \phi_k$ the $k$th eigenvalue and eigenfunction as before, and define $\mathcal{C}_{i,j}$ as having $[s,k]$th element
	
	\begin{equation} \label{cDef}
		\scriptstyle
		\mathcal{C}_{i,j}[s, k] = \begin{cases}
			0, & \text{if } 0 \leq t_s < a_{i,j} \\
			\tilde{\phi}_k(m_{i,j}(t_s)) - \tilde{\phi}_k(m_{i,j}(a_{i,j})), & \text{if } a_{i,j} \leq t_s < d_{i,j} \\ 
			\left( \tilde{\phi}_{k}(m_{i,j}(d_{i,j})) - \tilde{\phi}_k(m_{i,j}(a_{i,j})) \right) exp(-\lambda_{i,j} (t_s - d_{i,j})), & \text{if } d_{i,j} \leq t_s \\
		\end{cases}
	\end{equation}
	
	\noindent for unknown recovery parameter $\lambda_{i,j} > 0$, detailed below, and where 
	
	\begin{equation} \label{mFunDef}
		m_{i,j}(t) = min \left\{ 1, \frac{t - a_{i,j}}{M_\text{max}} \right\}.
	\end{equation}
	
	for maximum day of hospital stay $M_\text{max}$. The first line of (\ref{cDef}) allows $\mathcal{C}_{i,j}$ to have the same number of rows as $\mathcal{X}_i$, and prevents the admission from effecting hemoglobin for $t \leq a_{i,j}$. Ignoring the $m_{i,j}$ function for the moment, the second line determines the effect of hospitalization from admission to discharge, the subtraction enforces continuity at $a_{i,j}$. The first term on the third line is the final hospitalization time; when post-multiplied by $\bm{\beta}_{i,j}$ the final hospitalization effect is carried forwards. The exponential term decays to zero as time since discharge increases, so the patient gradually returns to their trend, $\mathcal{X}_i \bm{\alpha}_i$. No intercept is included in $\mathcal{C}_{i,j}$, as its inclusion would result in a discontinuity in hemoglobin at $a_{i,j}$. 
	
	The purpose of the $m_{i,j}$ function in (\ref{mFunDef}) is to provide additional flexibility over a short timescale, and to protect against highly uncertain estimation of the effects of long hospitalizations. Recall that the BSS basis requires input to be scaled to $[0,1]$, while, as described in Section \ref{dataSect}, the full time span of the study is two years. In order to allow a hospitalization of  6 days (the average hospitalization duration in our data, 0.8\% of the study duration) to substantially alter hemoglobin at the original timescale would require a large number of basis functions. Not only would this be computationally inefficient, it would allow for very highly variable estimation of the effects of the small number of long hospitalizations. The $m_{i,j}$ function allows much more flexibility over the length of most hospitalizations with a small number of basis functions. For hospitalizations longer than $M_\text{max}$, $m_{i,j}$ is constant, which implies a constant hospitalization effect under (\ref{cDef}). $M_\text{max}$ should therefore be chosen to be longer than the majority of hospitalizations. We use $M_\text{max} = 14$ days, as less than 7\% of hospitalizations exceed this length.
	
	We could use the same prior structure as (\ref{alpha0Prior}) and (\ref{alphaPrior}) for $\bm{\beta}_{i,j}$. However, the effects of hospitalization are likely to depend on a patient's hemoglobin; severely anemic patients are likely to receive blood transfusions and therefore experience a temporary hemoglobin increase, while patients with higher hemoglobin often experience hemoglobin losses due to surgery or other procedures. We therefore allow each $\beta_{i,j,k}, k = 1, \dots, b$ component to depend on the patient's hemoglobin at time of admission, 
	
	\begin{equation} \label{betaModel}
		\begin{split}
			&\beta_{i,j,k} | \eta_{k,0}, \eta_{k,1}, \omega^2 \overset{ind} {\sim} N(\eta_{k,0} +  \sqrt{\pi_k} T_i(a_{i,j}) \eta_{k,1}, \omega^2), \\
			& i = 1, \dots, n, \; j = 1, \dots, J_i, \; k = 1, \dots, b, \\
		\end{split}
	\end{equation}
	
	\noindent where $T_{i}(a_{i,j})$ is the true hemoglobin at time of admissios calculated according to (\ref{readmitResponse}). For $j > 1$ this includes the lingering effects of past hospitalizations.
	
	For computational ease we choose
	
	\begin{equation} \label{betaPrior}
		\begin{split}
			&\eta_{k,m} \overset{iid}{\sim} N(0, \sigma^2_{\eta,m}), m = 1, 2,\; k = 1, \dots b,  \\
			& \omega^2 \sim inv. Gamma(a_\omega, b_\omega), \\
			&\sigma^2_{\eta,m} \sim inv. Gamma(a_\eta, b_\eta), m = 1, 2. \\
		\end{split}
	\end{equation}.
	
	Because patient recovery rate, governed by $\lambda_{i,j}$, is likely to depend on that patient's overall health, we use a parametric random effects model. Let $\bm{q}_{i,j}$ be a vector of covariates for the $j$th hospitalization that are likely to effect recovery, such as age, and gender, see Table \ref{covariateTab} in the SM for the complete list. We also include the true hemoglobin at time of admission, $T_i(a_{i,j})$ and discharge, $T_i(d_{i,j})$, as well as an interaction between them. This allows each recovery to depend on how a patient's hemoglobin changed over the course of the hospitalization,so patients whose hemoglobin increased during hospitalization, likely due to a transfusion, can recover at a different rate than patients whose hemoglobin decreased. We then assume

	\begin{equation} \label{lambdaModel}
		\begin{split}
			& \lambda_{i,j} | \bm{\zeta}, \sigma^2_\lambda \overset{ind}{\sim} logNormal(\bm{q}_{i,j}^T \bm{\zeta}, \sigma^2_\lambda ), i = 1, \dots, n, \; j = 1, \dots, J_i, \\
			& \sigma^2_\lambda \sim inv. Gamma(a_\lambda, b_\lambda). \\
		\end{split}
	\end{equation}
	
	We use the BSS-ANOVA basis for $\bm{q}_{i,j}$. Let $z = 1, \dots, Z$ index the covariates included in each $\bm{q}_{i,j}$, and let $\bm{q}_{i,j,z}$ be the BSS expansion of covariate $z$, with corresponding vector of regression coefficients $\bm{\zeta}_z$. Then $\bm{q}_{i,j} = (\bm{q}_{i,j,1}^T, \dots, \bm{z}_{i,j,Z}^T)^T$ and $\bm{\zeta} = (\bm{\zeta}_{1}^T, \dots, \bm{\zeta}_Z^T)^T$, with the BSS formulation implying that 
	
	\begin{equation} \label{zetaPrior}
		\begin{split}
			& \zeta_{z,k} | \tau^2_{\zeta, z} \overset{iid}{\sim} N(0, \sigma^2_{\zeta, z}), z = 1, \dots, Z, \; k = 1, \dots, N_z \\
			& \tau^2_{\zeta, z} \overset{iid}{\sim} inv. Gamma(a_\zeta, b_\zeta), \\
		\end{split}
	\end{equation}
	
	\noindent where $N_z$ is the number of elements in the basis expansion for variable $z$. 
	
	Lastly, a note on dependence. Consider hospitalization $j^\prime$, and its attendant random effects $\bm{\beta}_{i,j^\prime}$ and $\lambda_{i,j^\prime}$, both of which depend on, and alter, true hemoglobin. Due to the first row of (\ref{cDef}), $\bm{\beta}_{i,j^\prime}$ only impacts $\bm{T}_i$ for $t > a_{i,j}$. by the third line of (\ref{cDef}), $\lambda_{i,j^\prime}$ only impacts $\bm{T}_i$ for $t > d_{i,j}$. Therefore the distribution of $\bm{\beta}_{i,j}$ depends on $\bm{\beta}_{i,j^\prime}$ and $\lambda_{i,j^\prime}$ only when $j > j^\prime$. Likewise, the distribution $\lambda_{i,j}$ depends on $\bm{\beta}_{i,j}$, but depends on $\lambda_{i,j^\prime}$ only when $j > j^\prime$. Therefore, none of the random effects depend on themselves or future events, and the model is well defined.
	
	\subsection{Admission and Discharge Event Models} \label{eventSect}
	We now have a model for a patient's overall hemoglobin trajectory, the impacts of hospitalization, and recovery from those impacts. However, (\ref{readmitResponse}) and its sequelae implicitly condition on hospitalizations being known \textit{a priori}. This is sufficient to fit a retrospective model, but our primary interest is predictive. Since hospitalizations are generally not known a year in advance, predicting the future from starting time $t^\star$ using (\ref{readmitResponse}) could use past hospitalizations, but cannot anticipate the rate or severity of future hospitalizations. This is obviously unsatisfactory, so we extend the model to include hospital admission and discharge processes. 
	
	Since both admits and discharges are likely to depend on hemoglobin, we use a JLM. Because patients can be in two states (at home or in the hospital) and neither state is absorbing, this is an alternating model \citep{hougaard1999multi}. Our admit/discharge data is an identical scenario to that described in \cite{lee2018semiparametric}'s analysis of alternating models. That work, however, models gap times and does not utilize a longitudinal covariate. 
	
	Because we are interested in hemoglobin over time, we need a model that allows easy use of longitudinal covariates. We also need to accomodate the alternating structure of the data. Considering only the admission process, a patient may be admitted multiple times, but is not at risk of hospital admission when in the hospital. Admission is therefore a recurrent model, with delayed (re)entry into the risk set during periods of hospitalization. The same holds for the discharge process. We therefore model both processes using the counting process representation described in \cite{kelly2000survival} for recurrent data, which allows delayed entry. Formally, define a patient's state at any given time as
	
	\begin{equation} \label{stateDef}
		A_i(t) = \begin{cases}
			1 & \text{ if patient $i$ is hospitalized at time $t$} \\
			0 & \text{ otherwise.}
		\end{cases}
	\end{equation} 
	
	We divide the full time time period of interest into short intervals $[t_{i,s}, t_{i,s+1})$ with length $\Delta_{i,s} = t_{i, s+1} - t_{i,s}$, where it is reasonable to assume that risk of readmission or discharge is approximately constant. We structure the retrospective data so that state changes only occur at the beginnings of time intervals, so each patient is in one state, and in one risk set, in each interval. 
	
	Let $Y_{i,A}(t_{i,s})$ be the number of events experienced by patient $i$ who is in state $A$ over the interval $[t_{i,s}, t_{i, s+1})$. We then assume that
	
	\begin{equation} \label{eventResponse} 
		Y_{i, A}(t_{i,s}) | A_{i}(t_{i,s}), h_{i,A}(t_{i,s}) \overset{ind}{\sim} \begin{cases} Poisson(\Delta_{i,s} h_{i,A}(t_{i,s}) ) \text{ if } A_{i}(t_{i,s}) = A \\
			0 \text{ with probability $1$ otherwise}
		\end{cases}
	\end{equation}
	
	\noindent where the point mass allows $f(Y_{i,A}(t_{i,s}))$ to be defined when $A_i(t_{i,s}) \neq A$. 
	
	Because the admission and discharge models are recurrent, there is information contained in the repeated event times for each patient that should not be discarded \citep{amorim2015modelling}. We therefore decompose each patient's hazard for each state change into an individual frailty random effect and a log-linear fixed effects model,
	
	\begin{equation} \label{hazDef}
		\begin{split}
			h_i(t_{i,s},A_i({t_{i,s}})) &= \rho_{i, A} exp(\bm{B}_{i,A}(t_{i,s})^T \bm{\psi}_A), \\
		\end{split}
	\end{equation}
	
	\noindent where $\bm{B}_{i,A}(t_{i,s})$ is a vector of covariates for patient $i$ in state $A$ at time $t_{i,s}$, $\bm{\psi}_A$ is a vector of regression coefficients, and $\rho_{i,A}$ is the frailty for patient $i$'s transitions from state $A$. While more complex models based on gap time are possible, this approach parsimoniously creates correlations within a patient's event history \citep{kelly2000survival}. We impose a gamma prior for the $\rho_{i,A}, A = 0, 1$,
	
	\begin{equation} \label{rhoDef}
		\begin{split}
			& \rho_{i,A} | \alpha_{\rho, A}, \beta_{\rho, A} \overset{iid}{\sim} Gamma (\alpha_{\rho, A}, \beta_{\rho, A}), \\
			& \alpha_{\rho, A} \sim Gamma(a_{\alpha, A}, b_{\alpha, A}), \; \beta_{\rho, A} \sim Gamma(a_{\beta, A}, b_{\beta, A}). \\
		\end{split}
	\end{equation}
	
	We allow the covariate vector $\bm{B}_{i,A}(t_{i,s})$ to depend on a patient's current hemoglobin, as well as the observation time and demographic variables, see Table \ref{covariateTab} in the SM for a complete list. As with $\bm{q}_{i,j}$, we use BSS decomposition for each covariate in $\bm{B}_{i,A}(t_{i,s})$. Let $d = 1, \dots, D_A$ index the covariates, $\bm{B}_{i,d,A}(t_{i,s})^T$ be the basis expansion for the $d$th covariate observed for patient $i$ at time $t_{i,s}$ for state $A$, and $\bm{\psi}_{d,A}$ the corresponding vector of regression coefficients. Then as in (\ref{zetaPrior}), we assume
	
	
	\begin{equation} \label{psiDef}
		\scriptstyle
		\begin{split}
			&\bm{\psi}_{d,A} \overset{ind}{\sim} N(\bm{0}, \nu^2_{d,A} I), d = 1, \dots, D_A, \; A = 0, 1, \\
			&\nu^2_{d,A} \overset{iid}{\sim} inv.Gamma(a_\nu, b_\nu), d = 1, \dots, D_A, \; A = 0, 1 \\
		\end{split}
	\end{equation}
	
	We allow the hazard to depend on a patient's current hemoglobin, as well as observation time, patient age and sex, and various health indicators, see the SM for a full list. For discharge hazard, we also include the length of the current impatient encounter, as with the hospitalization effect model this is scaled to maximum length of two weeks. 
	
	Because a patient must have an index encounter to be included in the data, the index encounter is fixed and known. Therefore, we must condition on the index admission. Formally, the conditional event likelihood is $f(Y_{i,0}(t_{i,s}), t_{i,s} \neq a_{i,\text{index}} | Y_{i,0}(a_{i,\text{index}}) ) = f(Y_{i,0}(t_{i,s}), s = 1, \dots, n_i)/f(Y_{i,0}(a_{i,\text{index}})$. Since the Poisson events are independent, the numerator is the product across observation times, and the index event drops out.  
	
	The index admission still generates $\bm{\beta}_{i, \text{index}}$ and $\lambda_{i, \text{index}}$, since this is associated with change of state. Since hospital discharges are always conditional on there having been a prior admission, the index discharge is included in the risk set. Hospital admissions from before the index admission are not treated as fixed and known, since the patient was at risk of a non-index admission during those times. 
	
	\subsection{Model Fitting and Prediction} \label{predSect}
	Because we are using a Bayesian approach, we fit the model using a Metropolis within Gibbs algorithm. This is detailed fully in the SM, however, it is worth describing here in overview. Closed form conjugate updates are available for many parameters; namely the variance parameters, population effects $\bm{\gamma}_0, \bm{\gamma}_k$, $\eta_{k,0}$, $\eta_{k,1}$ and $\bm{\zeta}$, and the individual frailties, $\rho_{i,A}$. Because of the Poisson likelihoods for the admission and discharge processes depend on true hemoglobin, no closed form is available for $\bm{\alpha}_{i}$, $\bm{\beta}_{i,j}$ or $\lambda_{i,j}$. Parameters $\alpha_{\rho,A}$ and $\beta_{\rho,A}$ also lack closed form updates. For these parameters we use Metropolis random walks with normal proposal distributions, the variances of which are tuned during burn-in for optimum acceptance rates. 
	
	We now describe a method for generating out of sample future hemoglobin predictions. Assume we have fit the model to training data, and have posterior distributions for $\bm{\gamma}, \bm{\eta}, \sigma^2$ etc. Suppose we have data on patient $i^\prime$, not in the training set, up until $t^\star$. We assume this includes states until $t^\star$, say $\bm{A}_{i^\prime}^\star$; the number of observed admissions, $J_{i^\prime}^\star$; and the admit and discharge times up until time $t^\star$. Observed hemoglobin values prior to $t^\star$, $\bm{H}_{i^\prime}^\star$ may also be available, but are not necessary to generate predictions. We assume that all demographic values for $\bm{z}_{i^\prime,0}, \bm{z}_{i^\prime}, \bm{q}_{i^\prime,j}$, $\bm{B}_{i^\prime, A}(t_{i^\prime, s})$ are observed. 
	
	We generate true hemoglobin curves conditional on the observed data as follows. For iterations $r = 1, \dots, R$, let $\sigma^{2,(r)}$, $\bm{\gamma}_0^{(r)}$, $\tau_0^{2,(r)}$, $\gamma_1^{(r)}, \dots, \gamma_{p-1}^{(r)}$, $\tau^{2,(r)}$ etc be a sample from the posterior of the fitted model. Run an MCMC chain of length $M$ for the individual parameters $\bm{\alpha}_{i^\prime}$, $\bm{\beta}_{i^\prime,j}, \rho_{i^\prime, A}, A = 0, 1$ and $\lambda_{i^\prime, j}, j = 1, \dots J_{i^\prime}^\star$ conditional on $\bm{H}_{i^\prime}^\star$, $\bm{A}_{i^\prime}^\star$ and the sampled posterior values. No assumptions are made about timepoints for $t > t^\star$. We now have a true hemoglobin trajectory, which is conditional on all observed data up until time $t^\star$. However, since the event data stops at $t^\star$, it does not account for future hospitalizations and is not yet a complete prediction. 
	
	To generate the event sequence and its effect on the trajectory, choose a small time interval, $\delta$, over which the transition hazard should be relatively constant; we use $\delta = 1$ day. Calculate the transition hazard $h_A(t^\star)$ according to (\ref{hazDef}), using $\bm{\psi}_A^{(r)}$, and $T_{i^\prime}(t^\star)$ calculated from (\ref{readmitResponse}) using individual random effects estimated above. Then the number of events in $[t^\star, t^\star + \delta)$ has a Poisson distribution with mean $\delta \times h_A(t^\star)$; therefore the time between events is exponential with mean $1/h_A(t^\star)$.
	
	To determine if there is a transition in $[t^\star, t^\star + \delta)$ draw $\epsilon \sim exp(1/h_A(t^\star))$. If $\epsilon > \delta$, then the patient's status does not change in $[t^\star, t^\star + \delta)$, so set $t^\star = t^\star + \delta$, and repeat the previous step. If $\epsilon < \delta$, then the patient's status changes at time $t^\star + \epsilon$. If $i^\prime$ was not previously hospitalized, they are admitted at time $t^\star + \epsilon$, so set $J_{i^\prime}^\star = J_{i^\prime}^\star + 1$, draw $\bm{\beta}_{i^\prime, J_{i^\prime}^\star}$ from (\ref{betaModel}) for the new admission, and calculate $\mathcal{C}_{i,J_{i^\prime}^\star}$ from (\ref{cDef}). If the patient was already in the hospital, they are discharged at $b_{i,J_{i^\prime}^\star} = t^\star + \epsilon$, so draw a new $\lambda_{i,J_{i^\prime}^\star}$ from (\ref{lambdaModel}). Add new rows to $\mathcal{X}_{i^\prime}$, $\mathcal{C}_{i^\prime,j}, j = 1, \dots J_{i^\prime}^\star$ for the new timepoint. Set $t^\star = t^\star + \epsilon$ and repeat until $t^\star$ reaches the end of the time period of interest. 
	
	This method produces a posterior sample of $R$ complete hemoglobin trajectories, admissions and discharges, which are conditionally independent given the fit model and prior observed data for patient $i^\prime$. Any quantity calculated from the distribution of these trajectories will account for additional uncertainty due to hospitalization.
	
	We implemented the model in R \citep{RSoftware}, for efficiency we encapsulated patients using custom C++ objects built in Rcpp and Rcpp Armadillo \citep{Rcpp, Armadillo}. Due to the large number of patients, we parallelized estimation of individual random effects using the foreach package \citep{calaway2015package}. 
	
	\section{Results} \label{resultsSect}
	We validate our model using five-fold cross validation, with patients randomized to folds. For each hold out set $f^\prime = 1, \dots, 5$, we use the methods detailed in Section \ref{predSect} and the SM to fit the model to the full two years of data for patients in training folds $f \neq f^\prime$. We then use the fit model to generate $R = 100$ trajectories for new patients $i^\prime$ in fold $f^\prime$, conditioned on hemoglobin and hospitalization data up until a given cutoff time $t^\star$ as defined in Section \ref{predSect}. We use cutoff times  $t^\star = 365, 395, 425, \dots, 695, 730$, which allows us to investigate how the amount of historical data impact prediction we provide results. Therefore all results presented in this section are based on out-of-sample predictions. In particular, predictions from $t^\star = 365$ days demonstrate how well the model performs for a patient when predicting from index discharge, and are analagous to predictions for a new patient being discharged from the ICU.  
	
\begin{figure}[h!]
		\centering
		\includegraphics[width=\textwidth]{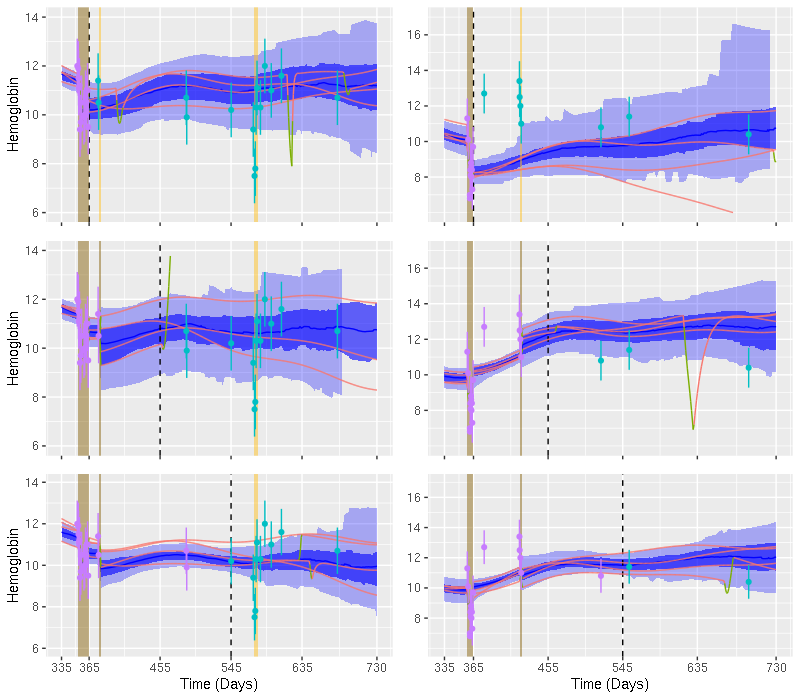}
		\caption{\label{predPatientPlot} \footnotesize 0-day (first row), 90-day (second row), 180-day (third row) predictions for the patients in Figure \ref{rawHbFig}; time of prediction is shown by the vertical dashed line. Hemoglobin observations are shown as dots with $+/- 2$se error bars, se is estimated as the posterior median, hospitalizations are shown using vertical rectangles, dark to the left of the prediction line, light to the right. Hemoglobin and hospitalizations to the left of the prediction line are conditioned on to generate predictions. The posterior distribution of true hemoglobin, $\bm{T}_i$ is shown by 90\% and 50\% pointwise central credible sets (light and dark bands, respectively), with the posterior median in the center. The tan and green lines show randomly sampled simulated hemoglobin trajectories, with tan indicating that the patient is at home and green that the patient is hospitalized.}
	\end{figure}

	Figure \ref{predPatientPlot} shows predictions for the two patients from Figure \ref{rawHbFig}, calculated at index discharge and three, six months thereafter. The patient on the left suffers only a very slight decrease in hemoglobin due to their index encounter, and the 0-day predictor in the first row predicts the rest of their recovery quite well, except for the lowest observations during the hospitalization around 210 days. The impact of hemoglobin observations on uncertainty is clearly illustrated by the decrease in uncertainty between the day-90 and day-180 predictions caused by the two observations around day 110. The patient on the right recovers quickly from their index encounter, while the day-0 predictor guesses a slower recovery; however the day-90 predictor captures the behavior in the hospitalization around 60 days, and continues to perform well for the remaining times, as does the day-180 predictor. 
	
	It is not reasonable to calculate population level results such as area under the curve (AUC), calibration or mean absolute deviation (MAD) every day because most days have only a few observations. We therefore aggregate these quantities over 30 day intervals; $[365, 395)$, $[395, 425)$, $\dots$, $[635, 695)$, $[395, 730)$ days post discharge. The final interval is an extra five days long, which avoids a short ``rump" interval. 
	
\begin{figure}[h!]
		\centering
		\includegraphics[width=\textwidth]{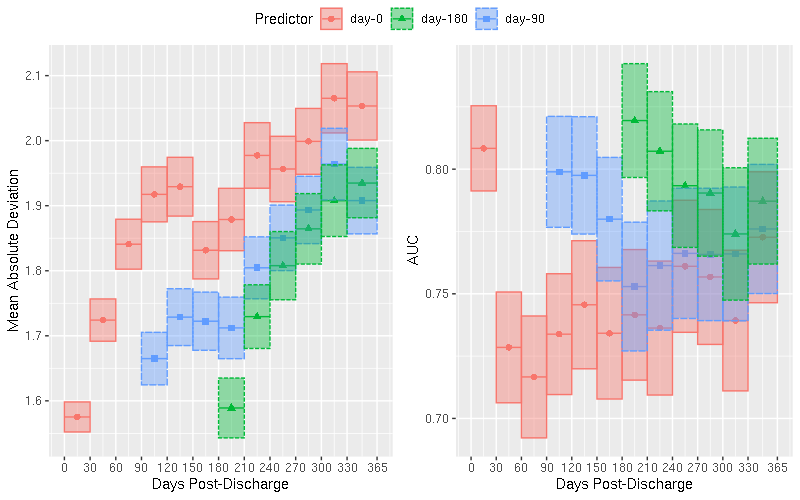}
		\caption{ \label{FinalCombPlot} \footnotesize Left: Mean Absolute Deviation (MAD) by Time, Data Available to Predictor. Error bars are $+/- 2$ standard errors. Right: Area Under the Curve by Time, Data Available to Predictor with 95\% confidence intervals calculated according to \cite{delong1988comparing}. Values of MAD/AUC are not shown for times prior to prediction time}. 
	\end{figure}

	Figure \ref{FinalCombPlot} shows the MAD of predicted hemoglobin, and the AUC for predicted recovery. MAD is defined as the absolute deviation between each observed hemoglobin value and the posterior median for the patient's predicted true hemoglobin on the day of observation, averaged across all observations in that interval. We define a patient as recovered if they have at least one observed hemoglobin value that exceeds the WHO sex-based anemia threshold (13 mg/dL for men, 12 mg/dL for women) in a time interval. We use the proportion of generated trajectories that exceed the recovery threshold in that same interval as a predictor. We truncate trajectories at time of death. This conditions on time of death, which is unknown at time of prediction. However, not doing so allows the patient extra time to recover under the model when compared to reality, resulting in bias. Further, in practice any question about hemoglobin recovery is implicitly conditional on the patient surviving. Since our goal is to predict hemoglobin and anemia, not mortality, stopping evaluation at time of death is the most fair comparison.  
	
	Model performance improves for both MAD and AUC as more data becomes available, and for intervals closer to the prediction time. For MAD, the day-0 predictor starts out quite accurate, then decays fairly quickly. This is likely because for the first 30 days post-discharge hemoglobin is generally low due to the index event, making it easy to predict. Over 30 - 150 days however, patients are recovering at a variety of rates, which causes the 0-day predictor to perform less well. The day-90 predictor does significantly better than the day-0 predictor, indicating that the first 90 days of data make a large difference in predicting recovery rate. The 180-day predictor is broadly comparable to the 90-day predictor except for days 180 - 210, and both substantially outperform the 0-day predictor. This indicates that patient dynamics for six months or more post-index discharge are simply difficult to predict, due to unknown hospital admissions and recoveries. 
	
	The MAD is calculated using the median of marginal distribution of hemoglobin for each day, which averages across at-home and hospitalized states. Because hemoglobin is generally lower in the hospital than at home, this causes the median to be biased low as an estimator when a patient is at home, and biased high when the patient is in fact in the hospital. This simple uncertainty in state appears to drive a fairly large portion of the error, for the 0-day predictor for instance the MAD over the entire year of followup is 1.83 g/dL, while the MAD for only at-home observations is 1.74 g/dL and only in-hospital observations is 1.95 g/dL. Because patients are mostly not hospitalized, most trajectories are not hospitalized at most times, and thus the magnitude of the bias is higher for in-hospital observations than at-home observations. 
	
	Broadly the same holds for AUC, which is not surprising since the recovery criterion is a function of hemoglobin. However for the 0-day predictor, the AUC is high for the first 30 days, probably due probably to the same depressed hemoglobin effect that drives the superior MAD in the first 30 days, and then slowly decreases. Although the 90-day and 180-day predictors have a more uniform decline in AUC, as time passes they fall towards the performance of the 0-day predictor. This suggests that after four to six months, the effects of proximity and additional data become fairly unimportant. This is in some sense reassuring, since it means that predicting a year ahead from index discharge is as valid an undertaking as making that same prediction six months later. Importantly, the AUC of all predictors remains in excess of .7 uniformly, suggesting that the proportion of trajectories that exceed the recovery threshold is a viable predictor
	
\begin{figure}[h!]
		\centering
		\includegraphics[width=\textwidth]{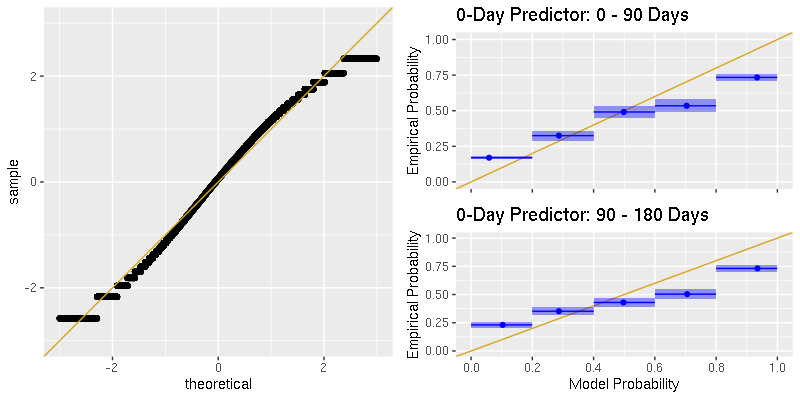}
		\caption{\label{qqPlot} \footnotesize Left: Normal Quantile-Quantile plot for the 0-day predictor over the first thirty days; the diagonal line is the identity $y = x$ line. The quantile for each observation is computed as (number of trajectories <= observation + 1/2)/(number of trajectories + 1/2), with the addition of 1/2 in the numerator and denominator to avoid undefined quantiles. Right Top: Calibration plot for the 0-day predictor over 0-30 days. Right Bottom: Calibration Plot for the 0-day predictor for 90 - 120 days. The confidence intervals are $2 \times s.e.$ where $s.e. = \sqrt{\hat{p}(1 - \hat{p})/n}$, where $\hat{p}$ is the proportion of patients who are observed to recover in the interval, and $n$ is the number of patients for whom recovery can be assessed. The dot in each interval is the mean of the model probabilities for individuals in that interval. The diagonal line is the $y = x$ identity line.}
	\end{figure}

	Figure \ref{FinalCombPlot} shows that our score function does well at predicting both hemoglobin and recovery, but the model may still be poorly calibrated to reality or systematically biased. We asses this using quantile-quantile plots and calibration plots, shown in Figure \ref{qqPlot}. We create the quantile-quantile plots by first estimating the posterior distribution of observed (not true) values at each observation time $t_s$ as $E_i(t_s)^{(r)} = H_i(t_s)^{(r)} + \epsilon_{i}(t_s)^{(r)}$, $\epsilon_{i}(t_s) \overset{iid}{\sim} N(0, \sigma^2)$ for each posterior realization $r = 1, \dots, R = 100$. We estimate $\sigma^2$ as the posterior median. Then we estimate the model quantile as $q_i(t_s) = \sum_r (E_i(t_s)^{(r)} \leq H_i(t_s))/R$ for each observed hemoglobin value $H_i(t_s)$. If the model is correct $q_i(t_s) \sim U(0, 1)$ by the probability integral transform, so $P(Z \leq q_i(t_s))$ for $Z \sim N(0, 1)$ gives the quantiles of a standard normal distribution. We then plot these observed quantiles against the theoretical quantiles of a standard normal distribution. We create calibration plots by binning the model probability of the estimated observation distribution $E_i(t_s)$ exceeding the recovery threshold, then calculating the empirical probability of recovery for all patients in each bin.
	
	The horizontal banding in the quantile-quantile plot is caused by the finite number of trajectories generated by the model, with only $R = 100$ iterations it is simply not possible to estimate any quantile below $q = .01$ or above $q = .99$. We therefore prune the plot at $+/- 3$ standard deviations in order to focus attention on performance in the center. Within this interval, the model quantiles are generally fairly close to the theoretical quantiles, although with slightly heavier tails. This to be expected, since the model generates both hospitalized and non-hospitalized states at any given timepoint, while any given observation will be either hospitalized or not. This will generally lead to the model estimate, which marginalizes across state, being somewhat overdispersed relative to the conditional observations. However, the performance indicates that the model predictions are not systematically too low or too high. Therefore although the marginal model predictions show greater variability than the conditional observed reality, they are still close enough to that reality to be useful. More importantly, they capture the genuine uncertainty in hemoglobin at time of prediction, as future hospitalizations would be unknown for a patient being discharged from the ICU. 
	
	The calibration plots suggest that the model is fairly well calibrated, although the model probability appears to be slightly higher than the empirical probability for empirical probabilities $\geq 0.5$. This is likely because each model trajectory is sampled once a day, and therefore has 30 chances to clear the recovery threshold per time period, while most patients have at most one or two observations in any 30-day period. This discrepancy causes the model to over-estimate recovery probability slightly. 
	
\begin{figure}[h!]
		\centering
		\includegraphics[width=\textwidth]{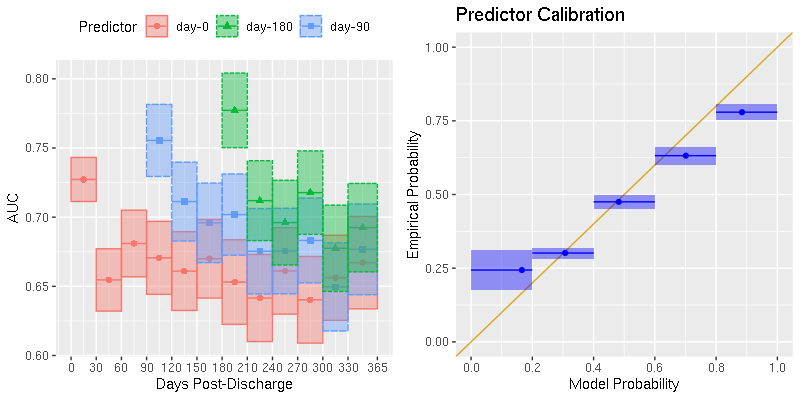}
		\caption{\label{hospFig} \footnotesize Left: Hospitalization AUC by time since discharge, predictor day. Right: Calibration of model to empirical probability for any hospitalization in the year following discharge. Empirical probabilities are shown with a 2 standard error confidence interval, where standard error is calculated as $\sqrt{\hat{p}(1 - \hat{p})/n}$ where $\hat{p}$ is the proportion of patients with a predicted hospitalization probability in each interval who are hospitalized in that interval. The diagonal line is the $y = x$ line indicating perfect calibration. The dots in each interval indicate the mean predicted probability for patients with predicted probabilities in the interval. The AUC of predicting any hospitalized status in the year following discharge is .70, 95\% CI (0.69, 0.72).}
	\end{figure}

	Because hospitalization drives so many of the behaviors exhibited in our data and model, we consider (\ref{eventResponse}) as a hospitalization predictor. We define a patient as hospitalized if they have any days in the hospital in a given 30 day interval. We use the proportion of patient trajectories that include at least one day in the hospital during that interval as a predictor. For patients who die, we only use trajectories up until time of death, for much the same reasons as discussed above Figure \ref{FinalCombPlot}. Figure \ref{hospFig} shows the per-month hospitalization AUC for the day-0, day-90  and day-180 predictors, and the calibration of the full year hospitalization predictor.
	
	
	Although the hospitalization is mostly included to allow more realistic hemoglobin prediction, the model still proves effective at predicting whether a patient will be hospitalized, both month-to-month and over the entire subsequent year. Our 30-day post-discharge AUC of .72 is competitive with the AUCs of the models summarized in Figure 6 of \cite{artetxe2018predictive}, which have an average AUC of .71 for 30-day readmission. This suggests that, although far from perfect, our model is doing as well as can be reasonably expected at predicting readmission. Further, our model's readmisison probabilities are very well calibrated to the true readmission chances, indicating the model's readmission probabilities are reliable. 
	
	\section{Conclusions} \label{concludeSect}
	Predictions of recovery from post-ICU anemia is a comparatively understudied issue in the literature. Although conceptually simple, the complex nature of patients' health over time, particularly the effects of (potentially repeated) hospital admissions, requires a fairly complex model to adequately address. By including stochastic readmission and discharge processes in our model, we are able to include both the effects of any hospitalizations prior to index discharge, as well as allow for the impacts of future hospitalizations on a patient's hemoglobin trajectory. Our model has shown itself capable of producing accurate predictions of whether a patient has recovered in a given time interval, and performs well at estimating actual hemoglobin levels. 
	
	However, because the model's predictions marginalize across the at home and hospitalized states, it generally overestimates hemoglobin when a patient is hospitalized and underestimates it when a patient is at home. Although in the validation this appears as bias due to the marginal predictor being compared to a conditional observation, this extra marginal uncertainty is correct. While we could compare a value observed on a non-hospitalized patient with trajectories that were not hospitalized at that time, this in effect conditions the validation on knowing the patient's state at time of prediction. While retrospectively this is possible, in practice for a new patient their future admissions and discharges will be unknown. The marginal predictor therefore gives the best view of that patient's potential range of future hemoglobins. 
	
	Although reasonable, our definition of being non-anemic is not ideal. For example a patient with hemoglobin readings of 13.5 mg/dL on day 10, 10.1 mg/dL on day 15 and 9.4 mg/dL on day 20 would be considered non-anemic for that month. However, because of the sharp decrease in hemoglobin, and clear development of anemia by the end of the month, this classification is of limited clinical use. We could classify each individual observation as anemic or not, and predict that instead of status over a month. This would render the definition extremely volatile in time, with patients shifting from being classified as anemic to non-anemic repeatedly over the course of days, which renders the classification itself less meaningful. This is simply a challenge with defining a binary state from a complex and evolving continuous longitudinal variable, and is difficult to resolve. However, the fairly accurate estimates of overall hemoglobin shown by the MAD panel of Figure \ref{FinalCombPlot}, indicates that we can reliably predict hemoglobin itself. Decisions about individual patients may then be made by referencing the predictions for that patient, such as those shown in Figure \ref{predPatientPlot}. 
	
	
	As previously described, we only use simulated trajectories up until time of death. If we include trajectories past time of death, the comparison of the model to the actual patient history is biased because the simulated trajectory has more time over which to generate high hemoglobin values that indicate recovery, or else be re-hospitalized. Further, any worry about a patient's future hemoglobin levels is implicitly conditioned on them being alive, which renders any bias due to not modeling patient death rather less important.
	
	Although not it's primary purpose, our model has shown itself fairly adept at predicting hospital admissions, and at least on this cohort, it compares well to existing predictors. Further work in this area is therefore very promising, particularly the expansion of the model to allow for multiple correlated longitudinal variables influencing the admission and discharge hazards. This could also extend to a more robust, time-varying approach to diagnosis codes or other variables that, due to limitations with the existing data, are only available at time of index encounter, and are therefore treated as constant throughout the model. Such work may also dovetail quite well with an expanded state space model, which includes death and potentially a more nuanced set of hospitalized states (ICU vs. non-ICU for example), which may be of substantial clinical interest.

	\textit{The authors report there are no competing interests to declare.} 
	\bibliography{Hbpaper1bibliography}
	
	\newpage
	
	\section{Supplemental Materials}
	
	\subsection{Hyperparameter Values} \label{hyperSect}
	The prior distributions described above are written in terms of general hyperparameters, rather than fixed values. This makes computing the form of updates simple (see Section \ref{updateSect} for details), however the values themselves must be specified in order to have a complete model. Table \ref{hyperTab} gives the values of all hyperparameters used in the model, as well as the distributions they appear in, their priors, and a brief description and notation of the relevant equations. 
	
	\begin{table}[H]
		\tiny
		\centering
		\begin{tabular}{l||l|l|l}
			Description & Model & Priors & Hyperparameters \\
			\hline \hline
			\begin{tabular}{l}Observed Hemoglobin  \\ see (\ref{responseModel}) and (\ref{readmitResponse}) \end{tabular}&  \begin{tabular}{l}$\bm{H}_i | \bm{T}_i, \sigma^2 \sim N(\bm{T}_i, \sigma^2 I)$ \\ $i = 1, \dots, n$ \end{tabular} & \begin{tabular}{l} $\sigma^2 \sim inv. Gamma(a_\sigma, b_\sigma)$ \end{tabular} & \begin{tabular}{l} $a_\sigma = 400001$, \\ $b_\sigma = 100000$ \end{tabular} \\
			\hline
			\begin{tabular}{l}Trend Intercept \\ see (\ref{alpha0Prior}) and (\ref{gammaPrior}) \end{tabular} & \begin{tabular}{l} $\alpha_{i,0} | \bm{\gamma}_0, \tau^2_0 \overset{ind}{\sim} N(\bm{z}_i^T \bm{\gamma}_0, \tau^2_0 I)$ \\ $i = 1, \dots, n$ \end{tabular} & \begin{tabular}{l}$\tau^2_0 \sim inv. Gamma(a_{\tau_0}, b_{\tau_0})$ \end{tabular} & \begin{tabular}{l} $a_{\tau_0} = 1$, \\ $b_{\tau_0} = 1$ \end{tabular} \\ 
			\hline
			\begin{tabular}{l} Trend Intercept Prior Mean \\ see (\ref{gammaPrior})\end{tabular} & \begin{tabular}{l}$\bm{\gamma}_0 | \sigma^2_{\gamma_0} \sim N(\bm{0}, \sigma^2_{\gamma_0} I)$ \end{tabular} & \begin{tabular}{l} $\sigma^2_{\gamma_0} \sim inv. Gamma(a_{\gamma_0}, b_{\gamma_0}) $ \end{tabular} & \begin{tabular}{l} $a_{\gamma_0} = 1$, \\ $b_{\gamma_0} = 1$ \end{tabular} \\
			\hline
			\begin{tabular}{l}Trend Comp. \\ see (\ref{alphaPrior}) and (\ref{gammaPrior}) \end{tabular} & \begin{tabular}{l} $\alpha_{i,k} | \bm{\gamma}_k, \tau^2 \overset{ind}{\sim} N(\sqrt{\pi_k} \bm{z}_i^T \bm{\gamma}_k, \tau^2)$, \\ $k = 1, \dots, p - 1$ \end{tabular} & \begin{tabular}{l} $\tau^2 \sim inv. Gamma(a_\tau, b_\tau)$ \end{tabular} &  \begin{tabular}{l} $a_\tau = 1,$ \\ $b_\tau = 1$ \end{tabular} \\
			\hline
			\begin{tabular}{l} Trend Comp. Prior Mean \\ see (\ref{gammaPrior}) \end{tabular} & \begin{tabular}{l} $\bm{\gamma}_k | \sigma^2_\gamma \sim N(\bm{0}, \pi_k \sigma^2_\gamma I)$ \\ $k = 1, \dots, p - 1$ \end{tabular} & \begin{tabular}{l} $\sigma^2_\gamma \sim inv. Gamma(a_{\sigma_\gamma}, b_{\sigma_\gamma})$ \end{tabular} & \begin{tabular}{l} $a_{\sigma_\gamma} = 1,$ \\ $b_{\sigma_\gamma} = 1$ \end{tabular} \\	
			\hline
			\begin{tabular}{l} Admit Effect Comp. \\ see (\ref{betaModel}) and (\ref{betaPrior}) \end{tabular} & \begin{tabular}{l} $\beta_{i,j,k} | \eta_{k,0}, \eta_{k,1}, \omega^2 \overset{ind} {\sim} $ \\ $ \; \; \; \; N(\eta_{k,0} +  \sqrt{\pi_k} T_i(a_{i,j}) \eta_{k,1}, \omega^2),$ \\
				$i = 1, \dots, n, \; j = 1, \dots, J_i, \; k = 1, \dots, b$ \\ \end{tabular} & \begin{tabular}{l} $\omega^2 \sim inv. Gamma(a_\omega, b_\omega) $ \end{tabular} & \begin{tabular}{l} $a_\omega = 1,$ \\ $b_\omega = 1$ \end{tabular} \\
			\hline
			\begin{tabular}{l} Beta Comp. Prior Mean \\ see (\ref{betaPrior}) \end{tabular} & \begin{tabular}{l} $\eta_{k,m} \overset{iid}{\sim} N(0, \sigma^2_\eta),$ \\ $m = 1, 2,\; k = 1, \dots b$ \end{tabular} & \begin{tabular}{l} $\sigma^2_\eta \sim inv. Gamma(a_\eta, b_\eta)$ \end{tabular} & \begin{tabular}{l} $a_\eta = 1$, \\ $b_\eta = 1$ \end{tabular} \\
			\hline
			\begin{tabular}{l} Recovery Effect \\see (\ref{cDef}) and (\ref{lambdaModel}) \end{tabular} & \begin{tabular}{l} $\lambda_{i,j} | \bm{\zeta}, \sigma^2_\lambda \overset{ind}{\sim} logNormal(\bm{q}_{i,j}^T \bm{\zeta}, \sigma^2_\lambda )$, \\ $i = 1, \dots, n, \; j = 1, \dots, J_i$ \end{tabular} & \begin{tabular}{l} $\sigma^2_\lambda \sim inv. Gamma(a_\lambda, b_\lambda)$  \end{tabular} & \begin{tabular}{l} $a_\lambda = 10$, \\ $b_\lambda = 10$ \end{tabular} \\
			\hline
			\begin{tabular}{l} Recovery Effect \\ Mean Comp. \\ see (\ref{zetaPrior}) \end{tabular} & \begin{tabular}{l}  $\zeta_{z,k} | \tau^2_{\zeta, z} \overset{iid}{\sim} N(0, \sigma^2_{\zeta, z}),$ \\ $z = 1, \dots, Z, k = 1, \dots, N_z$ \end{tabular} & 
			\begin{tabular}{l} $\tau^2_z \overset{iid}{\sim} inv.Gamma(a_\zeta, b_\zeta),$ \\ $z = 1, \dots, N_z$ \end{tabular} & \begin{tabular}{l} $a_\zeta = 1$, \\ $b_\zeta = 1$ \end{tabular} \\
			\hline
			\begin{tabular}{l} Frailty $\alpha$, $\beta$ Parameters \\ see (\ref{stateDef}), (\ref{eventResponse}), (\ref{hazDef}) and (\ref{rhoDef}) \end{tabular} & \begin{tabular}{l} $\rho_{i, A} | \alpha_{\rho, A}, \beta_{\rho, A} \sim Gamma(a_{\alpha ,a}, b_{\alpha, A})$,  \\$i = 1, \dots, n, \; A = 0, 1$ \end{tabular} & \begin{tabular}{l} $\alpha_{\rho, A} \sim Gamma(a_{\alpha, A}, b_{\alpha, A})$, \\ $\beta_{\rho, A} \sim Gamma(a_{\beta, A}, b_{\beta, A})$ \\ $A = 0, 1$  \end{tabular} &  \begin{tabular}{l} $a_{\alpha, A} = 2$, \\ $b_{\alpha, A} = 1/2$, \\ $a_{\beta,A} = 2$, \\ $b_{\beta, A} = 1/2$, \\ $A = 0, 1$  \end{tabular} \\
			\hline
			\begin{tabular}{l} Hazard Comp. \\ see (\ref{stateDef}), (\ref{eventResponse}), (\ref{hazDef}) and (\ref{psiDef}) \end{tabular} & \begin{tabular}{l} $\bm{\psi}_{d,A} \overset{ind}{\sim} N(\bm{0}, \nu^2_{d,A} I),$ \\ $d = 1, \dots, D_A, \; A = 0, 1$ \end{tabular} & \begin{tabular}{l} $\nu^2_{d,A} \overset{iid}{\sim} inv.Gamma(a_\nu, b_\nu),$ \\ $d = 1, \dots, D_A, \; A = 0, 1$ \end{tabular} & \begin{tabular}{l} $a_\nu = 1$, \\ $b_\nu = 1$, \\ $A = 0, 1$ \end{tabular} \\
			\hline
			
		\end{tabular}
		\caption{\label{hyperTab} \footnotesize Hyperparameters used, listed by relevant model. The relevant equation numbers for each model are listed in the second line of each model description. The abbreviation `Comp.' is used for Component; note also that the distribution for the Admit Effect Components is broken over two lines for reasons of space. }
	\end{table}
	
	\subsection{Parameter Likelihoods and Updates} \label{updateSect}
	
	Here we describe how to construct an MCMC algorithm for the model described above via a Metropolis within Gibbs strategy. Since this MCMC is designed to sample from a known, retrospective cohort, we assume hospitalization admission and discharge times are known. Throughout this section we use $f()$ as a generic notation for density, with $f(\theta | \circ)$ denoting the full conditional of $\theta$. For Metropolis steps we use the notation $\star$ to denote the candidate value of a parameter and ${(m)}$ for the current value; for notational simplicity we exclude the ${(m)}$ superscript on parameters not being updated in that step.

	\subsubsection{$H_i(t)$}
	
	Given the individual random effects and error variance $\sigma^2$, missing $H_{i}(t)$ may be drawn directly from (\ref{readmitResponse}) using their current values.
	
	\subsubsection{$\bm{T}_i$ random effects, $\bm{\alpha}_i$, $\bm{\beta}_{i,j}$ and $\lambda_{i,j}$}
	
	The individual random effects, $\bm{\alpha}_i$, $\bm{\beta}_{i,j}$ and $\lambda_{i,j}, j = 1, \dots, J_i$ collectively determine the true hemoglobin vector $\bm{T}_i$, as described in (\ref{readmitResponse}) and (\ref{cDef}). These parameters therefore have complex dependencies with other parameters, as well as each other; though as explained in Sections \ref{hospSect} and \ref{eventSect}, no parameter depends on itself or the event that causes it. However, the observed data $\bm{H}_i$ and $\bm{Y}_i$, and the other parameters and random effects depend on any given random effect only through the true hemoglobin vector $\bm{T}_i$. 
	
	Therefore drawing a new random effect is equivalent to drawing a new $\bm{T}_i$ vector when it comes to determining the dependence of other parameters on that random effect. Because the Poisson-distributed event data $\bm{Y}_i$ depends on true hemoglobin through (\ref{hazDef}), no conjugate update is possible, so we use self-tuning random walks for all random effects, as described in Section \ref{predSect}. While the acceptance ratios differ, we can use a common strategy and notation in all cases; draw a candidate value, denoted by the $^\star$ superscript, calculate the true hemoglobin vector $\bm{T}_i^\star$ under that candidate value, and likewise calculate the true hemoglobin vector $\bm{T}_i^{(m)}$ under the current value. From these, the likelihoods of any quantities that depend on the true hemoglobin can be readily calculated.
	
	For notational convenience we condense the event likelihoods into a single expression $f(\bm{Y}_i | \bm{A}_i, \bm{T}_i, \bm{\psi}_A, \rho_{i,A}) = \prod_{s = 1}^{n_i} f(Y_{i}(t_s) | A_i(t_s), T_i(t_s), \bm{\psi}_{A_i(t_s)}, \rho_{i, A})$. As both admission and discharge hazards are impacted by true hemoglobin, there is no benefit to writing out the likelihood of both event distributions in full detail here.

	First consider $\alpha_{i,0}$. Using the above strategy and notation, and writing $\bm{\eta}$ for the full vector of $\eta$ parameters (see Section \ref{etaSect} for details on rigorously defining the joint distribution of $\bm{\beta}_{i,j}$ in terms of $\bm{\eta}$), the acceptance ratio reduces to
	
	\begin{equation*}
		\scriptstyle
		\alpha = min \left\{ 1, \frac { f(\alpha_{i,0}^\star | \bm{\gamma}_0, \tau^2_0) f(\bm{H}_i | \bm{T}_i^\star, \sigma^2) f(\bm{Y}_i | \bm{A}_i, \bm{T}_i^\star, \bm{\psi}_A, \rho_{i,A}) \left( \prod_{j = 1}^{J_i} f(\bm{\beta}_{i,j} | \bm{T}_i^\star, \bm{\eta}, \omega^2) f(\lambda_{i,j} | \bm{T}_i^\star, \bm{\zeta}, \sigma^2_\lambda) \right) }  { f(\alpha_{i,0}^{(m)} | \bm{\gamma}_0, \tau^2_0)  f(\bm{H}_i | \bm{T}_i^{(m)}, \sigma^2) f(\bm{Y}_i | \bm{A}_i, \bm{T}_i^{(m)}, \bm{\psi}_A, \rho_{i,A}) \left( \prod_{j = 1}^{J_i} f(\bm{\beta}_{i,j} | \bm{T}_i^{(m)}, \bm{\eta}, \omega^2) f(\lambda_{i,j} | \bm{T}_i^{(m)}, \bm{\zeta}, \sigma^2_\lambda) \right) }  \right\}. 
	\end{equation*}
	
	Note that all hospitalization effects $\bm{\beta}_{i,j}$ and $\lambda_{i,j}, j = 1, \dots J_i$ are impacted by $\alpha_{i,0}$, since it is a parameter of the trend model, which effects the true hemoglobin at all times.

	Now consider updating $\alpha_{i,k}$, for component $k = 1, \dots, p - 1$. Because $\bm{\alpha}_i$ can be high dimensional, we update it a component at a time via random walk. Using the same notation as before, the acceptance ratio reduces to 
	
	\begin{equation*}
		\scriptstyle
		\alpha = min \left\{ 1, \frac { f(\alpha_{i,k}^\star | \bm{\gamma}_k, \tau^2) f(\bm{H}_i | \bm{T}_i^\star, \sigma^2) f(\bm{Y}_i | \bm{A}_i, \bm{T}_i^\star, \bm{\psi}_A, \rho_{i,A}) \left( \prod_{j = 1}^{J_i} f(\bm{\beta}_{i,j} | \bm{T}_i^\star, \bm{\eta}, \omega^2) f(\lambda_{i,j} | \bm{T}_i^\star, \bm{\zeta}, \sigma^2_\lambda) \right) }  { f(\alpha_{i,k}^{(m)} | \bm{\gamma}_k, \tau^2)  f(\bm{H}_i | \bm{T}_i^{(m)}, \sigma^2) f(\bm{Y}_i | \bm{A}_i, \bm{T}_i^{(m)}, \bm{\psi}_A, \rho_{i,A}) \left( \prod_{j = 1}^{J_i} f(\bm{\beta}_{i,j} | \bm{T}_i^{(m)}, \bm{\eta}, \omega^2) f(\lambda_{i,j} | \bm{T}_i^{(m)}, \bm{\zeta}, \sigma^2_\lambda) \right) }  \right\}
	\end{equation*}
	
	As before, the effects for all hospitalizations $j = 1, \dots, J_i$ are effected when updating $\alpha_{i,k}$. 
	
	The update for $\bm{\beta}_{i,j}$ is similar to the other individual random effects, and again we utilize a self-tuning normal random walk. As with the $\bm{\alpha}_i$, we draw each $\beta_{i,j,k}$ individually, which in our experience improves convergence. 
	
	Suppose we are updating component $k$ for admission $j^\prime$, for which we write $\beta_{i,j^\prime, k}$. Then the acceptance ratio reduces to 
	
	\begin{equation*}
		\scriptstyle
		\alpha = min \left\{ 1, \frac { f(\beta_{i,j^\prime, k}^\star | \eta_{k,0}, \eta_{k, 1}, T_i(a_{i,j^\prime}), \omega^2) f(\bm{H}_i | \bm{T}_i^\star, \sigma^2) f(\bm{Y}_i | \bm{A}_i, \bm{T}_i^\star, \bm{\psi}_A, \rho_{i,A}) f( \lambda_{i,j^\prime} | \bm{\zeta}, \bm{T}_i^\star, \sigma^2_\lambda)  \times p_{j^\prime}(\star) }  { f(\beta_{i,j^\prime, k}^{(m)} | \eta_{k,0}, \eta_{k, 1}, T_i(a_{i,j^\prime}), \omega^2)  f(\bm{H}_i | \bm{T}_i^{(m)}, \sigma^2) f(\bm{Y}_i | \bm{A}_i, \bm{T}_i^{(m)}, \bm{\psi}_A, \rho_{i,A}) f( \lambda_{i,j^\prime} | \bm{\zeta}, \bm{T}_i^{(m)}, \sigma^2_\lambda) \times p_{j^\prime}((m)) }  \right\} 
	\end{equation*}
	
	\noindent where
	
	\begin{equation*}
		\begin{split}
			&p_{j^\prime}(\star) =  \prod_{j = j^\prime + 1}^{J_i} f(\bm{\beta}_{i,j} | \bm{T}_i^\star, \bm{\eta}, \omega^2) f(\lambda_{i,j} | \bm{T}_i^\star, \bm{\zeta}, \sigma^2_\lambda)  \\
			&p_{j^\prime}((m)) =  \prod_{j = j^\prime + 1}^{J_i} f(\bm{\beta}_{i,j} | \bm{T}_i^{(m)}, \bm{\eta}, \omega^2) f(\lambda_{i,j} | \bm{T}_i^{(m)}, \bm{\zeta}, \sigma^2_\lambda) \\
		\end{split}
	\end{equation*}

	Note that the true hemoglobin at admission time, $T_i(a_{i,j^\prime})$ depends only on $\bm{\beta}_{i,j}, \lambda_{i,j}$ for $j < j^\prime$, and so it is the same under both the proposal and current value of $\beta_{i,j^\prime, k}$. Inpatient effects $\bm{\beta}_{i,j}, j > j^\prime$ and $\lambda_{i,j}, j \geq j^\prime$ are impacted by updating $\beta_{i,j^\prime, k}$, and so appear in the acceptance ratio, previous hospitalizations are not impacted by hospitalization $j^\prime$ and so fall out. 
	
	Lastly, consider updating $\lambda_{i,j^\prime}$. The acceptance ratio is then 
	
	\begin{equation*}
		\scriptstyle
		\alpha = min \left\{ 1, \frac { f(\lambda_{i,j^\prime}^\star | \bm{\zeta}, T_i(a_{i,j^\prime}), T_i(b_{i,j^\prime}), \sigma^2_\lambda) f(\bm{H}_i | \bm{T}_i^\star, \sigma^2) f(\bm{Y}_i | \bm{A}_i, \bm{T}_i^\star, \bm{\psi}_A, \rho_{i,A}) \left( \prod_{j = j^\prime + 1}^{J_i} f(\bm{\beta}_{i,j} | \bm{T}_i^\star, \bm{\eta}, \omega^2) f(\lambda_{i,j} | \bm{T}_i^\star, \bm{\zeta}, \sigma^2_\lambda) \right) }  { f(\lambda_{i,j^\prime}^{(m)} | \bm{\zeta}, T_i(a_{i,j^\prime}), T_i(b_{i,j^\prime}), \sigma^2_\lambda)  f(\bm{H}_i | \bm{T}_i^{(m)}, \sigma^2) f(\bm{Y}_i | \bm{A}_i, \bm{T}_i^{(m)}, \bm{\psi}_A, \rho_{i,A}) \left( \prod_{j = j^\prime + 1}^{J_i} f(\bm{\beta}_{i,j} | \bm{T}_i^{(m)}, \bm{\eta}, \omega^2) f(\lambda_{i,j} | \bm{T}_i^{(m)}, \bm{\zeta}, \sigma^2_\lambda) \right) }  \right\}.
	\end{equation*}
	
	As before, $T_i(a_{i,j^\prime})$ and $T_i(b_{i,j^\prime})$ do not depend on $\lambda_{i,j^\prime}$, and so are the same under both the current and candidate value. 
	
	\subsubsection{$\bm{\zeta}$}
	Due to the use of a lognormal recovery effect, $\lambda_{i,j}$, $\bm{\zeta}$ has a conjugate update. Define
	
	\begin{equation*}
		\bm{\lambda} = \begin{pmatrix}
			\lambda_{1,1} \\
			\vdots \\
			\lambda_{1, J_1} \\
			\lambda_{2, 1} \\
			\vdots \\
			\lambda_{n, J_n} \\
		\end{pmatrix}, \;
		\mathcal{Q} = \begin{pmatrix}
			\bm{q}_{1,1}^T \\
			\vdots \\
			\bm{q}_{1, J_1}^T \\
			\bm{q}_{2, 1}^T \\
			\vdots \\
			\bm{q}_{n, J_n}^T \\
		\end{pmatrix},
	\end{equation*}
	
	\noindent which allows us to write 
	
	\begin{equation*}
		\begin{split}
			&\bm{\zeta} | \circ \sim N(\bm{m}, V) \\
			& V = \left(I/\sigma^2_\zeta + \mathcal{Q}^T \mathcal{Q}/\sigma^2_\lambda \right)^{-1} \\
			& \bm{m} = V \times \mathcal{Q}^T(log(\bm{\lambda})/\sigma^2_\lambda \\.
		\end{split}
	\end{equation*}
	
	\subsubsection{$\sigma^2_\zeta$}
	$\sigma^2_\zeta$ also has a conjugate update, where
	
	\begin{equation*}
		\sigma^2_\zeta | \circ \sim inv. Gamma(a_\zeta + z/2, b_\zeta + \bm{\zeta}^T \bm{\zeta}/2). 
	\end{equation*}
	
	\subsubsection{$\bm{\gamma}_0$}
	Not surprisingly, $\bm{\gamma}_0$ has a standard normal-normal conjugate update. Some simple algebra shows that the full conditional is
	
	\begin{equation*}
		\begin{split}
			&\bm{\gamma}_0 | \circ \sim N(\bm{m}, V), \\
			&V = \left( I/\sigma^2_{\gamma_o} + n I/\tau^2_0 \right)^{-1}, \\
			&\bm{m} = V \left(\sum_{i = 1}^n \bm{z}_i \alpha_{i,0}/\tau^2_0 \right). \\
		\end{split}
	\end{equation*}
	
	\subsubsection{$\bm{\gamma}$}
	Each $\bm{\gamma}_k$ has a normal-normal update, and because of the independent priors each $\bm{\gamma}_k$ can be updated separately. It follows that
	
	\begin{equation*}
		\begin{split}
			&\bm{\gamma}_k | \circ \sim N(\bm{m}, V), \\
			&V = \left(  I/(\pi_k \sigma^2_\gamma) + \pi_k \sum_{i = 1}^n \bm{z}_i \bm{z}_i^T/\tau^2  \right)^{-1}, \\
			&\bm{\mu} = \sqrt{\pi_k} V \times \left(\sum_{i = 1}^N \alpha_{i,k} \bm{z}_i /\tau^2 \right).
		\end{split}
	\end{equation*}
	
	Although slightly odd in appearance, $V$ incorporates shrinkage sensibly. The prior precision $1/\sigma^2_\gamma$ is in effect reduced by $\pi_k$, and the amount of variability in the `data' $\bm{z}_i \bm{z}_i^T$ is also downscaled by $\pi_k$. Thus as $k$ increases and $\pi_k$ decreases, the posterior variance must also decrease since $\sigma^2_\gamma, \tau^2$ and $\bm{z}_i \bm{z}_i^T$ are identical across $k = 1, \dots, p - 1$. 
	
	\subsubsection{$\tau^2_0$ and $\tau^2$}
	Both $\tau^2_0$ and $\tau^2$ have standard inverse-Gamma conjugate update, although some care must be taken for the eigenvalue scaling. Let $\bm{\mu}_i = E(\bm{\alpha}_i |\bm{\eta}) = \mathcal{Z}_i \bm{\eta}$.
	
	Some straightforwards algebra shows that the full conditional of $\tau^2_0$ is 
	
	\begin{equation*}
		\tau^2_0 | \circ \sim inv. Gamma(a_{\tau_0} + n/2, b_{\tau^2_0} + (\mu_{i,0} - \alpha_{i,0})^2/2),
	\end{equation*}
	\noindent where $\mu_{i,0}$ is the element of $\bm{\mu}_i$ corresponding to the intercept $\alpha_{i,0}$.
	
	Similar algebra shows that the full conditional of $\tau^2$ is 
	
	\begin{equation*}
		\begin{split}
			&\tau^2 \sim inv. Gamma(a_\text{post}, b_\text{post}   ) \\
			&a_\text{post} = a_\tau +  n \times (p - 1)/2 \\
			&b_\text{post} = b_\tau + \sum_{i = 1}^n (\bm{\mu}_{i,(-0)} - \bm{\alpha}_{i,(-0)})^T D_{p - 1}(\bm{\pi}, (-0))^{-1} (\bm{\mu}_{i, (-0)} - \bm{\alpha}_{i, -(0)}) \\
		\end{split},
	\end{equation*}
	
	\noindent where $\bm{\mu}_{i, (-0)}$ is formed by dropping the intercept element of $\bm{\mu}_i$, and $D_{p - 1}(\bm{\pi}, (-0))$ is a diagonal matrix with $k$th diagonal element $\pi_k$. 
	
	\subsubsection{$\sigma^2_{\gamma_0}$}
	It follows from (\ref{gammaPrior}) and (\ref{alpha0Prior}) that
	
	\begin{equation*}
		\sigma^2_{\gamma_0} | \circ \sim inv. Gamma(a_{\gamma_0} + Z_0/2, b_{\gamma_0} + \bm{\gamma}_0^T \bm{\gamma}_0/2)
	\end{equation*}
	
	\noindent where $Z_0 = |\bm{z}_{i,0}|$ is the number of elements in $\bm{z}_{i,0}$. 
	
	\subsubsection{$\sigma^2_\gamma$}
	The update for $\sigma^2_\gamma$ is conjugate. From (\ref{gammaPrior}) and (\ref{alphaPrior}) it follows that
	
	\begin{equation*}
		\sigma^2_\gamma | \circ \sim inv Gamma \left(a_\gamma + Z \times (p - 1)/2, b_\gamma + \sum_{k = 1}^{p - 1} \pi_k^{-1} \bm{\gamma}_k^T \bm{\gamma}/2 \right),
	\end{equation*}
	
	\noindent where $Z = |\bm{z}_i|$ is the number of elements in each $\bm{z}_i$ vector.

	\subsubsection{$\omega^2$}
	
	As with $\tau^2_0$ and $\tau^2$, $\omega^2$ has a straightforward conjugate update. Some simple algebra shows that

	\begin{equation*}
		\begin{split}
			&\omega^2 | \circ \sim inv Gamma(a_\text{post}, b_\text{post} )\\
			&a_\text{post} = a_\omega +  d \sum_{i = 1}^n J_i/2 \\
			&b_\text{post} =  b_\omega + \sum_{i = 1}^n \sum_{j = 1}^{J_i} \sum_{k = 1}^d \left(\beta_{i,j,k} - (\eta_{k,0} + \sqrt{\pi_k} T_{i(a_{i,j})} \eta_{k,1}) \right)^2/2, \\
		\end{split}
	\end{equation*}

	\noindent where $d$ is the length of $\bm{\beta}_{i,j}$.
	
	\subsubsection{$\eta$ Parameters} \label{etaSect}
	The updates for the $\eta$ parameters that govern the mean of $\bm{\beta}_{i,j}$ are relatively straightforward. Because the draw is conjugate, and the conditionally independent prior specified in (\ref{betaPrior}) implies an easy to work with joint distribution, it is easiest to update all $\eta$ parameters at once. To that end, let $\bm{\eta} = (\eta_{0,0}, \eta_{0, 1}, \dots, \eta_{b, 0}, \eta_{b, 1})^T$. Then let $\mathcal{W}_{i,j}$ be a $b \times 2b$ matrix of the form
	
	\begin{equation*}
		\mathcal{W}_{i,j} = \begin{pmatrix}
			1 & \sqrt{\pi_0} T_i(a_{i,j}) & 0 & 0 & \dots & \dots & 0 & 0 \\
			0 & 0 & 1 & \sqrt{\pi_1} T_i(a_{i,j}) & \dots & \dots & 0 & 0 \\
			\vdots & \vdots & \vdots & \vdots & \ddots & \ddots & \vdots & \vdots \\
			0 & 0 & 0 & 0 & \dots & \dots & 1 & \sqrt{\pi_b} T_i(a_{i,j}) \\
		\end{pmatrix},
	\end{equation*}
	
	\noindent which allows us to write
	
	\begin{equation*}
		\begin{split}
			&\bm{\eta} | \sigma^2_\eta \sim N(\bm{0}, \sigma^2_\eta I) \\
			&\bm{\beta}_{i,j} | \bm{\eta}, \omega^2 \overset{ind}{\sim} N(\mathcal{W}_{i,j}, \omega^2 I), i = 1, \dots, n, \; j = 1, \dots, J_i, \\
		\end{split}
	\end{equation*}
	
	\noindent which is simply a restatement of (\ref{betaModel}) and (\ref{betaPrior}) explicitly in terms of the joint distributions of $\bm{\eta}$ and $\bm{\beta}_{i,j}$. From this it immediately follows that 
	
	\begin{equation*}
		\begin{split}
			&\bm{\eta} | \circ \sim N(\bm{m}, V) \\
			&V = \left(I/\sigma^2_\eta + \sum_{i = 1}^n \sum_{j = 1}^{J_i} \mathcal{W}_{i,j}^T \mathcal{W}_{i,j}/\omega^2 \right)^{-1}  \\
			&\bm{m} = V \times \sum_{i = 1}^n \sum_{j = 1}^{J_i} \mathcal{W}_{i,j}^T \bm{\beta}_{i,j}/\omega^2. \\
		\end{split}
	\end{equation*}

	\subsubsection{$\sigma^2_\eta$} 
	The update for $\sigma^2_\eta$ is an entirely straightforwards conjugate draw
	
	\begin{equation*}
		\begin{split}
			\sigma^2_\eta | \circ \sim inv. Gamma(a_\eta + 2b/2, b_\eta + \sum_{k = 0}^b \sum_{m = 1}^2 \eta_{k,m}^2/2),
		\end{split}
	\end{equation*}
	
	\noindent recalling that $b$ is the number of $\bm{\beta}$ components; to emphasize the mean of each $\beta$ component is determined by two $\eta$ components, we do not reduce the fraction $2b/2$. 
	
	\subsubsection{$\psi$ parameters}
	We update the $\bm{\psi}$ vectors using a covariate-by-covariate self-tuning random walk. That is, because in the BSS-ANOVA basis, binary covariates such as gender are represented by two basis functions, we draw a proposal for both the corresponding regression coefficients at once. Because we only use a linear effect for continuous covariates in the hazard model, these covariates only get a single basis element, and so are drawn individually. 
	
	Suppose we are updating $\bm{\psi}_{d,A}$ for covariate $d, d = 1, \dots, D_A$ and state $A = 0, 1$. The acceptance ratio reduces to 
	
	\begin{equation*}
		\alpha = min \left\{ 1, \frac{ \prod_{i = 1}^n f( \bm{Y}_{i,A} | \bm{\psi}_{d,A}^\star, \bm{\psi}_{-(d), A}^{(m)}, \bm{T}_i, \rho_{i,A} ) f(\bm{\psi}_d^\star | \nu_{d,A})  } { \prod_{i = 1}^n f( \bm{Y}_{i,A} | \bm{\psi}_A^{(m)}, \bm{T}_i, \rho_{i,A} ) f(\bm{\psi}_d^{(m)} | \nu_{d,A}) } \right\},
	\end{equation*} 
	
	\noindent where $\bm{\psi}_{{-d}, A}$ is the vector of all $\psi$ components for state $A$ except for component $d$. Note that because each $\bm{Y}_i$ in the denominator depends solely on the current values, the above equation simplifies the notation to $\bm{\psi}_A^{(m)}$, rather than $\bm{\psi}_{d, A}^{(m)}, \bm{\psi}_{-(d), A}$. 
	
	\subsubsection{$\nu^2_{d,A}$ parameters}
	Each of the $\nu^2_{d,A}$ parameters is a simple conjugate update. Recall that $\bm{\psi}_{d,A}$ has length $l_{d,A}$, so
	
	\begin{equation*}
		\nu^2_{d,A} | \circ \sim inv Gamma(a_\nu + l_{d_A}/2, b_\nu + \sum_{k = 1}^{l_{d,A}} \psi_{k,d,A}^2/2).
	\end{equation*}
	
	\subsubsection{$\rho_{i,A}$}
	The update for each $\rho_{i,A}$ is conjugate, with full conditional 
	
	\begin{equation*}
		\begin{split}
			&\rho_{i,A | \circ} \sim Gamma(a_{post}, b_{post}) \\
			&a_{post} = \alpha_{\rho, A} + \sum_{s = 1}^{n_i} Y_{i}(t_s) \\
			&b_{post} = \beta_{\rho, A} + \sum_{s = 1}^{n_i} \Delta_{i,s} exp(\bm{B}_{i,A}(t_i,s) \bm{\psi}_{A}).
		\end{split}
	\end{equation*}
	
	\subsubsection{$\alpha_{\rho, A}$ and $\beta_{\rho, A}$}
	The updates for  $\alpha_{\rho, A}$ and $\beta_{\rho, A}$ are very similar, so we group them together. Neither has a conjugate update, so we use separate self-tuning random walks with normal proposals for each parameter. Since $\alpha_{\rho, A}$ and $\beta_{\rho, B}$ are restricted to be positive, we automatically reject all proposals $\leq 0$. For $\alpha_{\rho, A}$ the acceptance ratio reduces to
	
	\begin{equation*}
		\alpha = min \left\{1, \frac{ f(\alpha_{\rho, A}^\star | a_{\alpha, A}, b_{\alpha, A}) \prod_{i = 1}^n f( \rho_{i, A} | \alpha_{\rho, A}^\star, \beta_{\rho, A} ) }{ f(\alpha_{\rho, A}^{(m)} | a_{\alpha, A}, b_{\alpha, A}) \prod_{i = 1}^n f( \rho_{i, A} | \alpha_{\rho, A}^{(m)}, \beta_{\rho, A} ) } \right\}, A = 0, 1,
	\end{equation*}
	
	\noindent similarly, the acceptance ratio for $\beta_{\rho, A}$ is
	
	\begin{equation*}
		\alpha = min \left\{1, \frac{ f(\beta_{\rho, A}^{\star} | a_{\beta, A}, b_{\beta, A}) \prod_{i = 1}^n f( \rho_{i, A} | \alpha_{\rho, A}, \beta_{\rho, A}^\star ) }{ f(\beta_{\rho, A}^{(m)} | a_{\beta, A}, b_{\beta, A}) \prod_{i = 1}^n f( \rho_{i, A} | \alpha_{\rho, A}, \beta_{\rho, A}^{(m)} ) } \right\}, A = 0, 1. 
	\end{equation*}

	\section{Covariates and Scaling}
	
	Because of the use of the BSS basis, each continuous covariate included in the model must be scaled to $[0,1]$. We generally do this by a simple transformation. Let $x$ by any covariate. Then for bounds $m_k < M_k$, let $\tilde{x}_{i,k}(t_s) = max \{m_k,  min \{\bar{x}_{i,k}(t_s), M_k  \}  \}$, where $\bar{x}_{i,k}(t_s)$ is the observed value. Then the final covariate value input in the model is
	
	\begin{equation*}
		\begin{split}
			x_{i,k}(t_s) = \frac{\tilde{x}_{i,k}(t_s - m_k)}{M_k - m_k}
		\end{split}
	\end{equation*}
	
	Note that this maps all $\tilde{x}_{i,k}(t_s) < m_k$ to 0, and all $\tilde{x}_{i,k}(t_s) > M_k$ to 1. By choosing $m_k, M_k$ close or equal to the data minimum and maximum the effects of truncation will be very small. Table \ref{covariateTab} shows $m_j, M_j$ for each covariate, the units of measure for that variable, as well as which portions of the model the covariate impacts. Categorical variables, such as gender, race and presence/absence of diagnosis codes are also listed. As described above (\ref{zetaPrior}) these use a different basis under BSS-ANOVA, and so are not scaled to $[0,1]$; the entries for $m, M$ for these variables are therefore blank. 
	
	\begin{table}[H]
		\footnotesize
		\centering
		\begin{tabular}{l|lccl}
			Variable & Unit & $m$ & $M$ & Models \\
			\hline
			Hemoglobin & mg/dL & 2.2 & 20 & $\bm{\beta}_{i,j}$, $\lambda_{i,j}$, admit, discharge hazards \\
			Study Time & Days & 0 & 730 & $\mathcal{X}_i$, admit, discharge hazards \\
			Inpatient Time & Days & 0 & 14 & $\mathcal{C}_{i,j}$, discharge hazard \\
			\hline 
			\hline
			Intercept & Unitless & & & $\alpha_{i,k}$, $\alpha_{i,0}$, $\lambda_{i,j}$, admit, discharge hazards \\  
			Age & Years & 16 & 100 & $\lambda_{i,j}$, $\alpha_{i,0}$, admit, discharge hazards \\
			Gender & Male/Female  & & & $\lambda_{i,j}$, $\alpha_{i,0}$, admit, discharge hazards \\
			Charlson Score & Unitless & 0 & 19 & $\lambda_{i,j}$, $\alpha_{i,0}$, admit, discharge hazards \\
			Supplemental Iron & Yes/No & & & $\lambda_{i,j}$, $\alpha_{i,0}$, admit, discharge hazards \\
			Renal Disease & Yes/No & & & $\alpha_{i,0}$, admit, discharge hazards \\
			Metastatic Solid Tumor & Yes/No & & & $\alpha_{i,0}$, admit, discharge hazards \\
			Other Cancer Dx & Yes/No & & & $\alpha_{i,0}$, admit, discharge hazards \\
			Index Surgical Admit & Yes/No & & & $\alpha_{i,0}$, admit, discharge hazards \\
			Race & See Caption & & & $\alpha_{i,0}$, admit, discharge hazards \\
			Index Day 1 SOFA Score & Unitless & 0 & 20 &  $\alpha_{i,0}$, admit, discharge hazards \\
			Hospital RBC Day 1 & Units & 0 & 107 &  $\alpha_{i,0}$, admit, discharge hazards \\
			Index ICU Days & Days & 0 & 30 &  $\alpha_{i,0}$, admit, discharge hazards \\ 
			Index Hospital Days & Days & 0 & 50 &  $\alpha_{i,0}$, admit, discharge hazards \\ 
			Total Hospital Duration & Days & 0 & 83 &  $\alpha_{i,0}$, admit, discharge hazards \\ 
			Total Emergency Department Duration & Days & 0 & 62 &  $\alpha_{i,0}$, admit, discharge hazards \\
			Total EMG Duration & Days & 0 & 50 &  $\alpha_{i,0}$, admit, discharge hazards \\ 
			Total Hospital Days & Days & 0 & 9 & $\alpha_{i,0}$, admit, discharge hazards \\
		\end{tabular}
		\caption{\label{covariateTab} \footnotesize The model includes six racial/ethnic categories, African, African American, Indian/Alaskan Native, Asian, White and Unknown/Other. All racial or ethnic groups that do not fall into one of the first five categories are mapped to Unknown/Other. All variables above the double horizontal bar change in time, so are calculated at each timepoint when the model is run. All variables below are considered baseline variables which do not change over the course of the study. Total Hospital Duration, Total ED Duration and total EMG duration are the total number of days prior to the index encounter the patient has spent in each of those departments. }
	\end{table}
	
	A notation of $\lambda_{i,j}$ in the Models column indicates the variable is included in the distribution of $\lambda_{i,j}$ as specified in (\ref{lambdaModel}); $\alpha_{i,0}$ means the variable is included in the distribution of $\alpha_{i,0}$ in (\ref{alpha0Prior}), $\alpha_{i,k}$ that the variable enters into (\ref{alphaPrior}) for $k = 1, \dots, p - 1$, $C_{i,j}$ means that the variable is used in the calculation of (\ref{cDef}), $\bm{\beta}_{i,j}$ means the variable determines the distribution of each $\beta_{i,j,k}$ as given in (\ref{betaModel}). Note that time in study determines $\mathcal{X}_i$ as described in (\ref{responseModel})

\end{document}